\documentclass[12pt]{article}
\pdfoutput=1

\usepackage{clay} 
\usepackage{amsmath}
\usepackage{amsfonts}
\usepackage{hyperref}
\numberwithin{equation}{section}
\usepackage{graphicx,color,subfig}
\usepackage{mciteplus}
\usepackage{skak}
\usepackage[nosort]{cite}
\DeclareFontFamily{OT1}{pzc}{}
\DeclareFontShape{OT1}{pzc}{m}{it}{<-> s * [1.10] pzcmi7t}{}
\DeclareMathAlphabet{\mathpzc}{OT1}{pzc}{m}{it}

\def\be#1\ee{\begin{align}#1\end{align}}

\newcommand{\bZ}{\mathbb{Z}}
\newcommand{\Z}{\mathbb{Z}}

\def\be{\begin{equation}}
\def\ee{\end{equation}}
\def\bea{\begin{eqnarray}}
\def\eea{\end{eqnarray}}
\def\ie{\begin{equation}\begin{aligned}}
\def\fe{\end{aligned}\end{equation}}

\begin{document}
\date{October, 2019}

\institution{Chicago}{\centerline{${}^{1}$Kadanoff Center for Theoretical Physics \& Enrico Fermi Institute, University of Chicago}}
\institution{IAS}{\centerline{${}^{2}$School of Natural Sciences, Institute for Advanced Study}}
\institution{Simons}{\centerline{${}^{3}$Simons Center for Geometry and Physics, SUNY}}
\institution{Rutgers}{\centerline{${}^{4}$NHETC and Department of Physics and Astronomy, Rutgers University}}

\title{Decorated $\mathbb{Z}_{2}$ Symmetry Defects and\\ Their Time-Reversal Anomalies}

\authors{Clay C\'{o}rdova,\worksat{\Chicago, \IAS}\footnote{e-mail: {\tt clayc@uchicago.edu}}
Kantaro Ohmori,\worksat{\Simons,\IAS}\footnote{e-mail: {\tt komori@scgp.stonybrook.edu}}
 Shu-Heng Shao,\worksat{\IAS}\footnote{e-mail: {\tt shao@ias.edu}}
Fei Yan\worksat{\Rutgers}\footnote{e-mail: {\tt fei.yan@physics.rutgers.edu}}
}

\abstract{We discuss an isomorphism between the possible anomalies of $(d+1)$-dimensional quantum field theories with $\mathbb{Z}_{2}$ unitary global symmetry, and those of $d$-dimensional quantum field theories with time-reversal symmetry $\mathsf{T}$.  This correspondence is an instance of symmetry defect decoration.  The worldvolume of a $\mathbb{Z}_{2}$ symmetry defect is naturally invariant under $\mathsf{T},$ and bulk $\mathbb{Z}_{2}$ anomalies descend to $\mathsf{T}$ anomalies on these defects.  We illustrate this correspondence in detail for $(1+1)d$ bosonic systems where the bulk $\mathbb{Z}_{2}$ anomaly leads to a Kramers degeneracy in the symmetry defect Hilbert space, and exhibit examples. We also discuss $(1+1)d$ fermion systems protected by $\mathbb{Z}_{2}$ global symmetry where interactions lead to a $\mathbb{Z}_{8}$ classification of anomalies.  Under the correspondence, this is directly related to the $\mathbb{Z}_{8}$ classification of $(0+1)d$ fermions protected by $\mathsf{T}$.  Finally, we consider $(3+1)d$ bosonic systems with  $\mathbb{Z}_{2}$ symmetry where the possible anomalies are classified by $\mathbb{Z}_{2}\times \mathbb{Z}_{2}$.  We construct topological field theories realizing these anomalies and show that their associated symmetry defects support anyons that can be either fermions or Kramers doublets. }

\maketitle

\setcounter{tocdepth}{3}
\tableofcontents

\section{Introduction}

Global symmetries and anomalies are crucial tools in the analysis of quantum field theory.  In its most elementary incarnation, global symmetry organizes Hilbert spaces into representations and gives rise to selection rules.  Anomalies are more subtle characteristic invariants of symmetry in quantum field theory and are constant along all continuous symmetry-preserving deformations of the theory, including scaling transformations of the renormalization group.  A rich class of phenomena are associated to systems with discrete global symmetry.  In this case the associated anomalies are also finite order and can be carried by gapped or gapless systems.  A precise understanding of the physics of anomalies is a central tool in many recent developments including the theory of topological insulators and superconductors, duality in $(2+1)d,$ and topological field theory.  

In this paper, we discuss an interesting connection between anomalies of $\mathbb{Z}_{2}$ unitary global symmetry in $d$ spacetime dimensions, and anomalies for antiunitary time-reversal ($\mathsf{T}$) symmetries in $(d-1)$ spacetime dimensions.  Since anomalies of $d$ dimensional theories are determined by inflow from $(d+1)$-dimensional SPTs, our discussion can also be interpreted as a connection between SPTs.   Specifically, there is an isomorphism:
\begin{equation}\label{bossmith}
\mathbb{Z}_{2}~\text{anomalies in}~d ~\text{dimensions} \longleftrightarrow \mathsf{T}~\text{anomalies in}~(d-1) ~\text{dimensions}~.
\end{equation}
Mathematically, SPT phases (including interactions) are classified by cobordism theory  \cite{KTalk1, Kapustin:2014tfa, Kapustin:2014gma, Kapustin:2014dxa, KTalk, Freed:2016rqq, Gaiotto:2017zba, Yonekura:2018ufj}. The precise mathematical relationship behind \eqref{bossmith} is sometimes referred to as a {\it Smith isomorphism} \cite{bahri1987eta,GilkeyBook} (see also \cite{Kapustin:2014dxa, Tachikawa:2018njr}), which we review in Appendix \ref{Appsmith}.  

Our main goal in this paper is to explore the physical consequences of \eqref{bossmith} in low-dimensional field theory examples.  In particular, we illustrate this correspondence in a simple class of models described by $(1+1)d$ bosonic QFTs with $\mathbb{Z}_{2}$ global symmetry.  We argue that this symmetry is anomalous if and only if the $(0+1)d$ symmetry line has states that are related by a Kramers degeneracy.  We also discuss the extension to $(1+1)d$ fermionic systems, and give an example application in $(3+1)d$ bosonic systems.  

\subsection{$\mathbb{Z}_{2}$ Symmetry Defects and Time-Reversal Symmetry}\label{genres}

\begin{figure}[h]
\centering
\includegraphics[width =.3\textwidth]{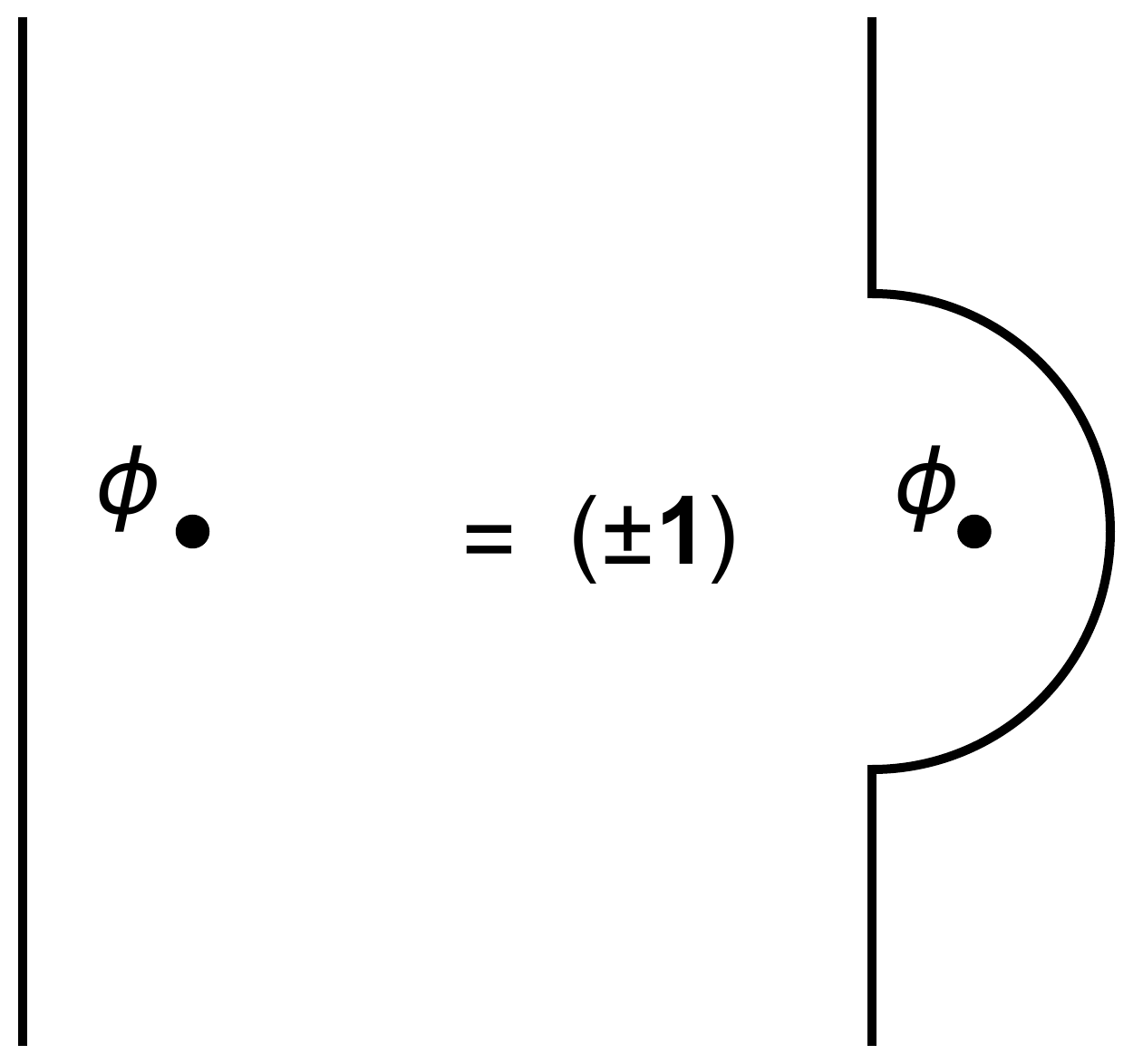}
\caption{The action of a $\bZ_2$ symmetry defect.  The defect is topological and small changes in its position do not modify correlation functions.  However, when the defect crosses a local operator $\phi$, the correlation changes by $\pm1$ depending on whether $\phi$ is even or odd under the action of the symmetry.}
\label{fig:0Z2}
\end{figure}

Physically, the isomorphism \eqref{bossmith} is mediated by symmetry defects.  In general in QFT, the abstract meaning of the symmetry, illustrated in Figure \ref{fig:0Z2}, is that there are codimension one topological operators that implement the $\mathbb{Z}_{2}$ symmetry action (see e.g.\ \cite{Gaiotto:2014kfa}). 
Equivalently one can view the presence of symmetry defects in correlation functions as coupling the QFT to a background $\mathbb{Z}_{2}$ gauge field.\footnote{Mathematically, such a gauge field is a cocycle $a\in Z^{1}(M,\bZ_{2})$, where $M$ is the $d$-dimensional spacetime.  The symmetry defect is then Poincar\'{e} dual to the cocycle $a$. The associated gauge equivalence class is $[a]\in H^{1}(M,\bZ_{2})$. }

The key observation behind the isomorphism \eqref{bossmith} is then that in a $d$-dimensional QFT with $\mathbb{Z}_{2}$ global symmetry, the symmetry defect is a $(d-1)$-dimensional quantum system with $\mathsf{T}$ symmetry.  This can be understood geometrically.  The analog of a symmetry defect for $\mathsf{T}$ is a locus where the orientation of spacetime is reversed.  Equivalently, this means that a QFT has $\mathsf{T}$-symmetry if it can be formulated on non-orientable manifolds.  In general, symmetry defects are defined on oriented manifolds, and changing the orientation leads to a defect for the inverse symmetry group element.  However, in the special case of $\mathbb{Z}_{2},$ the non-trivial group element is its own inverse and hence the defect is invariant under changing its orientation.   As argued above, this means that the defect theory has $\mathsf{T}$ symmetry.  (For more details see Appendix \ref{Appsmith}.)

\begin{figure}[h]
\centering
\includegraphics[width =.8\textwidth]{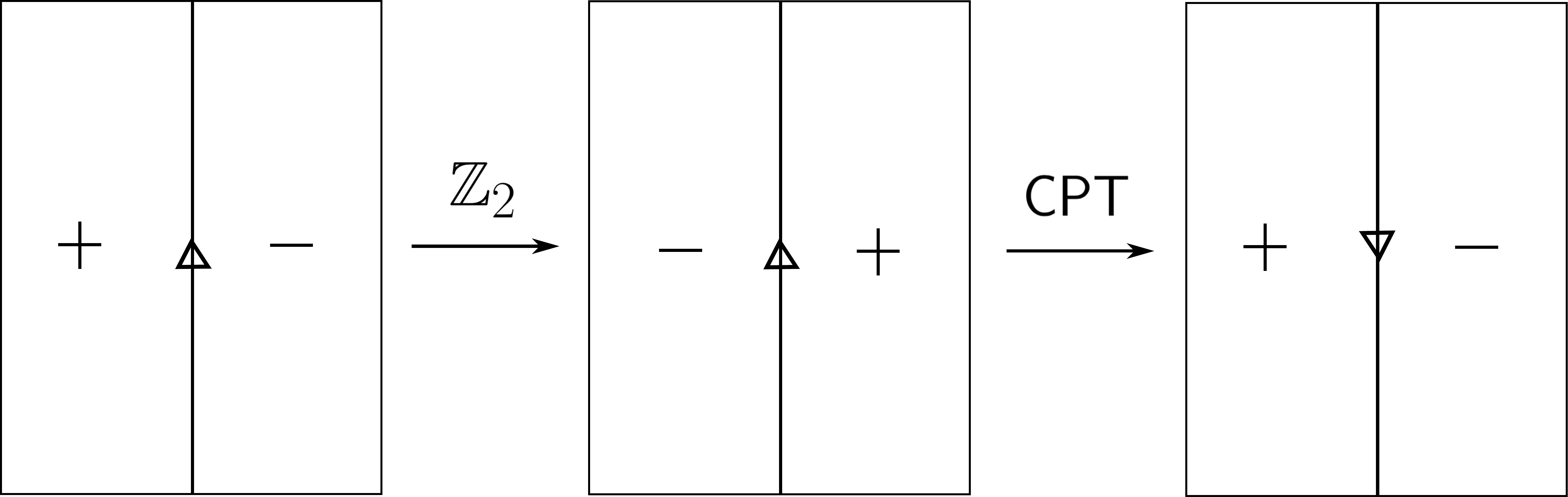}
\caption{The $\mathsf{T}$ symmetry on the symmetry defect in a spontaneously broken phase.  The defect separates two regions with degenerate ground states labelled by $\pm$.  The action of the $\mathbb{Z}_{2}$ symmetry exchanges the vacua, and combining with $\mathsf{CPT}$ leads to a symmetry of the original configuration that changes the worldvolume orientation of the defect.}
\label{fig:CPT}
\end{figure}

Another useful point of view can be obtained in the phase where the $\mathbb{Z}_{2}$ symmetry is spontaneously broken, illustrated in Figure \ref{fig:CPT}.  In that case the $\mathbb{Z}_{2}$ symmetry defect is realized as a domain wall separating two distinct vacua.  A global $\mathbb{Z}_{2}$ transformation does not preserve this configuration since it exchanges the vacua.  However if we combine this $\mathbb{Z}_{2}$ with the universal $\mathsf{CPT}$ symmetry, which acts in Euclidean signature as a rotation by $\pi$, we indeed find a symmetry of the original domain wall configuration.  As this combined action reverses the orientation of the $\mathbb{Z}_{2}$ symmetry defect, it descends to a $\mathsf{T}$ symmetry on its worldvolume. 

Thus, given any QFT with unitary $\mathbb{Z}_{2}$ global symmetry, the symmetry defect yields a theory in one lower dimension with antiunitary $\mathsf{T}$ symmetry.  Moreover, if the latter is anomalous, necessarily so is the former.  In fact, because \eqref{bossmith} is an isomorphism, all non-trivial $\mathbb{Z}_{2}$ anomalies of the bulk can be detected in this way.  Using inflow, the same arguments apply to the associated SPTs characterizing the anomalies.

The idea that the worldvolume theory of symmetry defects can be used to encode bulk anomalies is widely applied in the condensed matter literature \cite{Chen_2014}.  Here we see that the isomorphism \eqref{bossmith} gives an instance of this idea, where the $\mathbb{Z}_{2}$ symmetry defects are decorated theories with anomalous $\mathsf{T}$ symmetry.  The novelty however, is that the $\mathsf{T}$ symmetry on the defect is not a symmetry of the bulk.

One sharp way to realize the results of this correspondence is by circle compactification.  Consider a $d$-dimensional theory on a geometry $S^{1}\times \mathcal{L}$ for some $(d-1)$ manifold $\mathcal{L}$, and let a $\mathbb{Z}_{2}$ symmetry defect wrap $\mathcal{L}$ (and hence be located at a point in the $S^{1}$).  Equivalently, this means that there is a non-trivial $\mathbb{Z}_{2}$ holonomy around the $S^{1}$.   We can describe this configuration as an effective $(d-1)-$dimensional QFT.  According to the analysis above, this QFT has $\mathsf{T}$ symmetry, with a $\mathsf{T}$ anomaly that completely encodes the original $\mathbb{Z}_{2}$ anomaly of the $d$-dimensional theory.  In particular, thinking of $\mathcal{L}$ as extended along time, this means that the defect Hilbert space $\mathcal{H}_{\mathcal{L}}$ of the original theory (again with $\mathcal{L}$ localized at a point in a spatial circle) carries $\mathsf{T}$ symmetry.

\subsection{Implications for $(1+1)d$ QFT }

While the analysis above (and its generalization to spin theories discussed below) apply in arbitrary spacetime dimension, one focus of this paper is to understand in detail the map between $(1+1)d$ theories with $\mathbb{Z}_{2}$ symmetry, and quantum mechanics models with $\mathsf{T}$ symmetry.  

\begin{figure}[h!]
\centering
\includegraphics[width=.5\textwidth]{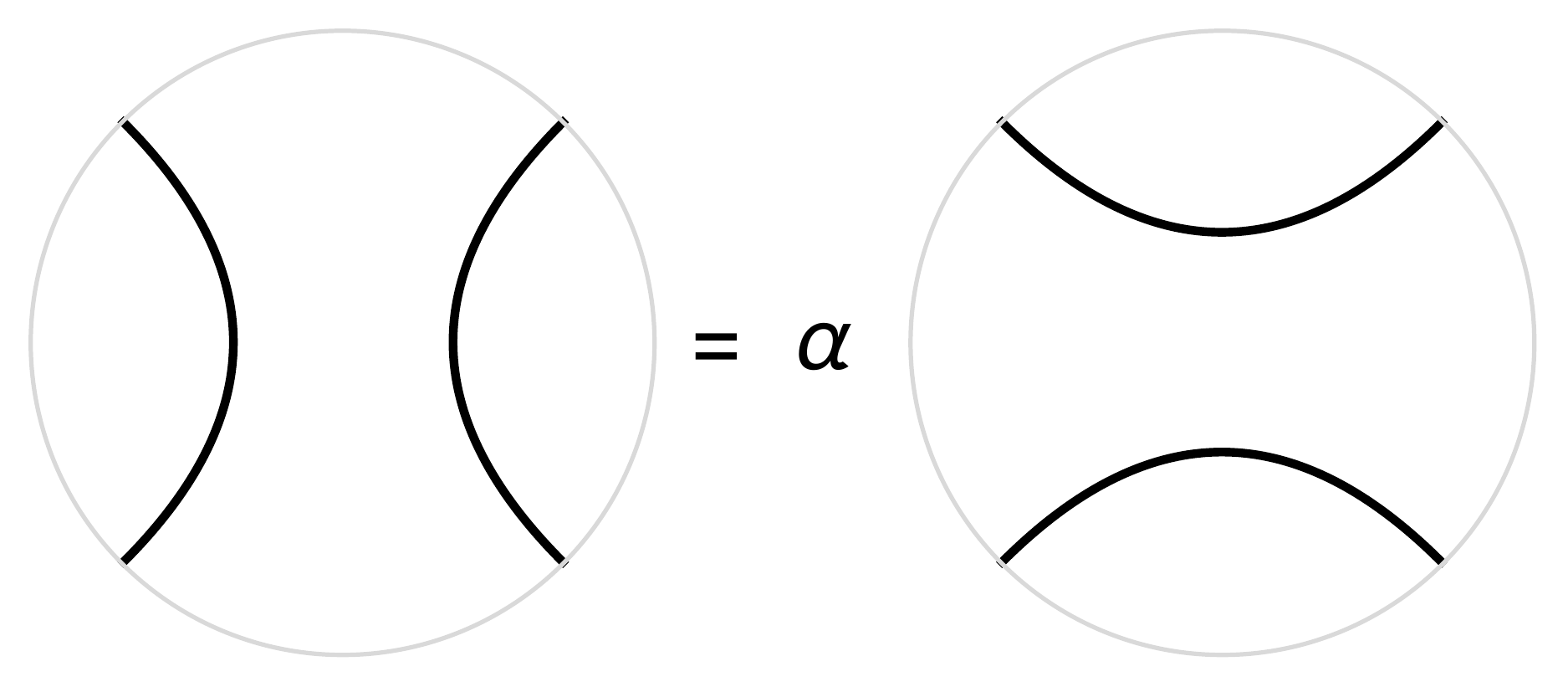}
\caption{The crossing relation of a $\mathbb{Z}_2$ line $\cal L$ (shown in black) in a local patch of the $(1+1)d$ geometry.  
The lines can be recombined at the cost of a phase $\alpha=\pm1$.  The $\mathbb{Z}_{2}$ symmetry is anomalous (i.e.\ there is an obstruction to orbifolding), if and only if $\alpha=-1$.}\label{fig:crossing}
\end{figure}

In the $(1+1)d$ case, the geometry of the symmetry defects and the map \eqref{bossmith} is particularly simple.  The $\mathbb{Z}_{2}$ symmetry is represented by a topological line $\mathcal{L}$ in the theory.  The general properties of these lines have been analyzed in \cite{Verlinde:1988sn, Petkova:2000ip,Fuchs:2002cm,Fuchs:2003id,Fuchs:2004dz,Fuchs:2004xi,Fjelstad:2005ua,Davydov:2010rm, Bhardwaj:2017xup,Chang:2018iay}.  For (bosonic) $(1+1)d$ QFTs with $\mathbb{Z}_{2}$ symmetry there is a unique possible non-trivial 't Hooft anomaly.\footnote{This anomaly is characterized by inflow from the $(2+1)d$ SPT  with classical action
\begin{equation}
\exp\left(\pi i \int_{N} a \cup a \cup a\right)~,
\end{equation}
where again $a\in H^{1}(N,\mathbb{Z}_{2})$ is the class representing the $\mathbb{Z}_{2}$ gauge field. Alternatively, we can view this SPT as a characterizing the non-trivial element in the group cohomology, $H^{3}(\mathbb{Z}_{2}, U(1))\cong \mathbb{Z}_{2}$.}  Following the general analysis in \cite{Bhardwaj:2017xup,Chang:2018iay,Lin:2019kpn}, it is easy to realize that this anomaly controls the crossing relation of the symmetry defects illustrated in Figure \ref{fig:crossing}.  

We use these results to independently argue that the line $\mathcal{L}$ has a $\mathsf{T}$ symmetry which is anomalous if and only if the $(1+1)d$ theory has an anomalous $\mathbb{Z}_{2}$ symmetry.  Since $\mathcal{L}$ is one-dimensional, this is an anomaly in quantum mechanics, i.e.\ a projective representation of the global symmetry.  For $\mathsf{T}$ symmetry this means that the system has an anomaly if and only if, on the Hilbert space:
\begin{equation}
\mathsf{T}^{2}=-1~.
\end{equation}
In other words, the Hilbert space consists of Kramers doublets.  We directly investigate defect Hilbert spaces, $\mathcal{H}_{\mathcal{L}}$ in a variety of examples of $(1+1)d$ CFTs and verify this conclusion.

\subsection{Generalization to Spin Systems}

Although much of our discussion concerns  bosonic theories.  The correspondence \eqref{bossmith} also holds for fermionic systems.  In this case, our system also has a $(-1)^{F}$ symmetry and we must say more about the algebra of the $\mathsf{T}$ symmetry in the correspondence.  For spin systems, there are two universal possibilities labelled $\text{pin}^{\pm}$ according to the implied structure group on spacetime:\footnote{In certain theories with additional global symmetry there can be other possible algebras obeyed by $\mathsf{T}$ see e.g. \cite{Cordova:2017kue}, however these do not play a role in the correspondence \eqref{bossmith}. }
\begin{equation}
\text{pin}^{+}\Longrightarrow \mathsf{T}^{2}=(-1)^{F}~, \hspace{.5in}\text{pin}^{-}\Longrightarrow\mathsf{T}^{2}=1~.
\end{equation}
As we review in Appendix \ref{Appsmith}, the precise map between $\mathbb{Z}_{2}$ anomalies in fermionic theories and $\mathsf{T}$ anomalies holds when the latter is part of a $\text{pin}^{-}$ symmetry group. 

In particular in Section \ref{fermsec}, we apply these ideas to systems of free fermions in $(1+1)d$.  There, as we review, the effect of interactions reduce the anomaly to a $\mathbb{Z}_{8}$ effect \cite{PhysRevB.81.134509,PhysRevB.85.245132,Qi_2013,PhysRevB.88.064507,PhysRevB.89.201113}.  The correspondence \eqref{bossmith} then implies the sharp result that the $(0+1)d$ fermion systems studied in \cite{Fidkowski_2011, PhysRevB.81.134509} also have mod eight periodicity.  Indeed, as we review, the ground state degeneracy of eight fermions with $\text{pin}^{-}$ symmetry can be removed by a $\mathsf{T}$-invariant interaction.

\subsection{Further Extensions}

Although we have focused above on $(1+1)d$ QFTs with $\mathbb{Z}_{2}$ symmetry, the correspondence \eqref{bossmith} holds for general dimensions.  As an illustration of this, in Section \ref{hdsec}, we consider $(3+1)d$ bosonic systems with $\mathbb{Z}_{2}$ unitary global symmetry.  These theories have anomalies classified by the group $\mathbb{Z}_{2}\times \mathbb{Z}_{2}$ and have recently been discussed in \cite{fidkowski2019disentangling}.  As predicted by the correspondence \eqref{bossmith}, the classification of $\mathsf{T}$ anomalies in $(2+1)d$ is also $\mathbb{Z}_{2}\times \mathbb{Z}_{2}$ \cite{Kapustin:2014tfa}.

We construct discrete gauge theories in $(3+1)d$ carrying these anomalies and explain how the $(2+1)d$ symmetry defect is decorated by an appropriate topological field theory with anomalous $\mathsf{T}$ symmetry where the lines can be either fermions or Kramers doublets.

\emph{Note added: This work is submitted in coordination with \cite{Zohar}, which contains partially overlapping results. We thank the authors of \cite{Zohar} for sharing their notes with us.}

\section{Bosonic $\mathbb{Z}_2$ Symmetry in $(1+1)d$}\label{bosonic}

In this section we investigate the correspondence \eqref{bossmith} in the simplest non-trivial class of examples:  $(1+1)d$ systems with $\mathbb{Z}_{2}$ global symmetry.  We show that the defect Hilbert space ${\cal H}_{\cal L}$ for an anomalous $\bZ_2$ line has Kramers degeneracy due to a time-reversal anomaly. 

\subsection{Symmetry Lines and the Defect Hilbert Space}

We start by reviewing some basic properties of the topological defect lines $\cal L$ implementing a unitary $\bZ_2$ global symmetry in $(1+1)d$.  A more detailed discussion can be found in \cite{Lin:2019kpn}.  Related ideas have been extensively discussed in \cite{Verlinde:1988sn, Petkova:2000ip,Fuchs:2002cm,Fuchs:2003id,Fuchs:2004dz,Fuchs:2004xi,Fjelstad:2005ua,Davydov:2010rm, Bhardwaj:2017xup,Chang:2018iay}.

A global symmetry in $(1+1)d$ is implemented by  a topological defect line $\cal L$.  The defining property of a topological defect line is that all physical observables such as correlation functions are invariant under small deformation of the locus of $\cal L$.  When the topological line is swept past a local operator $\phi(x)$, the correlation function is changed by the symmetry action of $\phi(x)$. For example, as we sweep a $\bZ_2$ line past a $\bZ_2$ even/odd local operator $\phi(x)$, the correlation function changes by $\pm1$. (See Figure \ref{fig:0Z2}.)

\begin{figure}[h!]
\centering
\includegraphics[width=.5\textwidth]{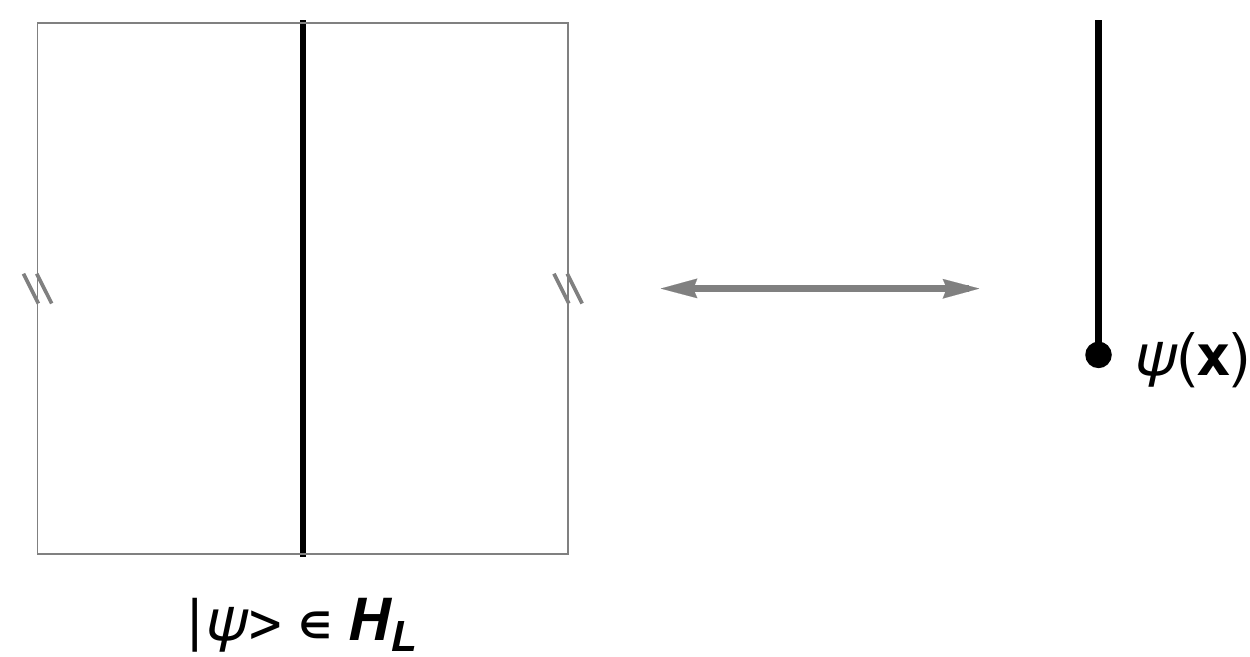}
\caption{The defect Hilbert space ${\cal H}_{\cal L}$ of a $\bZ_2$ line quantized on a circle $S^1$.  A state in the defect Hilbert space is mapped to an operator living at the end of the $\bZ_2$ line via the operator-state correspondence.}
\label{fig:HL}
\end{figure}

Consider the theory on a cylinder $S^1 \times \mathbb{R}$ with $\cal L$ running along the time $\mathbb{R}$ direction  (see Figure \ref{fig:HL}).   The topological line $\cal L$ intersects with the spatial $S^1$, and therefore modifies the quantization by a twisted periodic boundary condition. This defines a {\it defect Hilbert space} denoted by ${\cal H}_{\cal L}$.\footnote{When the $\bZ_2$ is non-anomalous, the $\bZ_2$ even sector ${\cal H}_{\cal L}^+$ of the defect Hilbert space ${\cal H}_{\cal L}$ before gauging is the twisted sector of the orbifold theory.}  This is a different Hilbert space than the Hilbert space $\cal H$ without the $\bZ_2$ line.

In the special case of conformal field theory,  a state in the usual Hilbert space $\cal H$ is mapped to a local operator via the operator-state correspondence.  A defect Hilbert space state $|\psi\rangle\in {\cal H}_{\cal L}$, on the other hand, is mapped to a non-local point-like operator living at the end of the $\bZ_2$ line. Since the topological line commutes with the stress tensor,  the states in the defect Hilbert space ${\cal H}_{\cal L}$  are organized into representations of the left and right Virasoro algebras. In particular,  the defect Hilbert space states can be diagonalized to have definite conformal weights $(h,\bar h)$.  In general, a state in ${\cal H}$ has integer spin $s=h-\bar h$, whereas as we will see below, a state in the  defect Hilbert space ${\cal H}_{\cal L}$ may have fractional spin  \cite{Chang:2018iay,Lin:2019kpn} (see also \cite{Hung:2013cda}).

We can also consider the insertion of multiple symmetry lines extended along time. This defines a multi-defect Hilbert space ${\cal H}_{\cal L L{\cdots}L}$.  However, in the special case of $\bZ_2$, we can fuse the symmetry lines pairwise to the trivial line.  In particular this means that the multi-defect Hilbert spaces with an even number of $\mathbb{Z}_{2}$ symmetry lines are all isomorphic to the Hilbert space $\cal H$ of local operators, while those with an odd number of lines are all isomorphic to the defect Hilbert space ${\cal H}_{\cal L}$. This also means that the defect Hilbert space ${\cal H}_{\cal LL}$, which via the operator-state map are the operators living \textit{on} the line, is isomorphic to the space $\cal H$ of ordinary local operators.

\subsection{$(1+1)d$ Bosonic $\bZ_2$ Anomaly}

The 't Hooft anomaly of  a $\bZ_2$ symmetry is encoded in a crossing relation of the topological symmetry lines.  Consider a local patch of a general correlation function as in the gray circle on the left of Figure \ref{fig:crossing}.  In this local patch we assume there are only two segments of $\bZ_2$ lines, without other local operator insertions.  Now imagine we replace the configuration in the local patch by the one on the right, without modifying the configuration outside the local gray patch.   Notice that to continuously interpolate from one figure to another, we must pass through a singular configuration where the lines cross.

We would like to compare the two correlation functions, before and after the replacement.  In a theory with a unique vacuum, the disc geometries in Figure \ref{fig:crossing} define states on the circle with defect lines inserted and both states have $h=\bar{h}=0$.  Therefore these two correlation functions can at most differ by a phase $\alpha$\cite{Chang:2018iay,Lin:2019kpn}.  Applying the same crossing rule twice, we conclude that $\alpha^2=1$. Hence $\alpha=\pm1$.

\begin{figure}[h!]
\centering
\includegraphics[width=.8\textwidth]{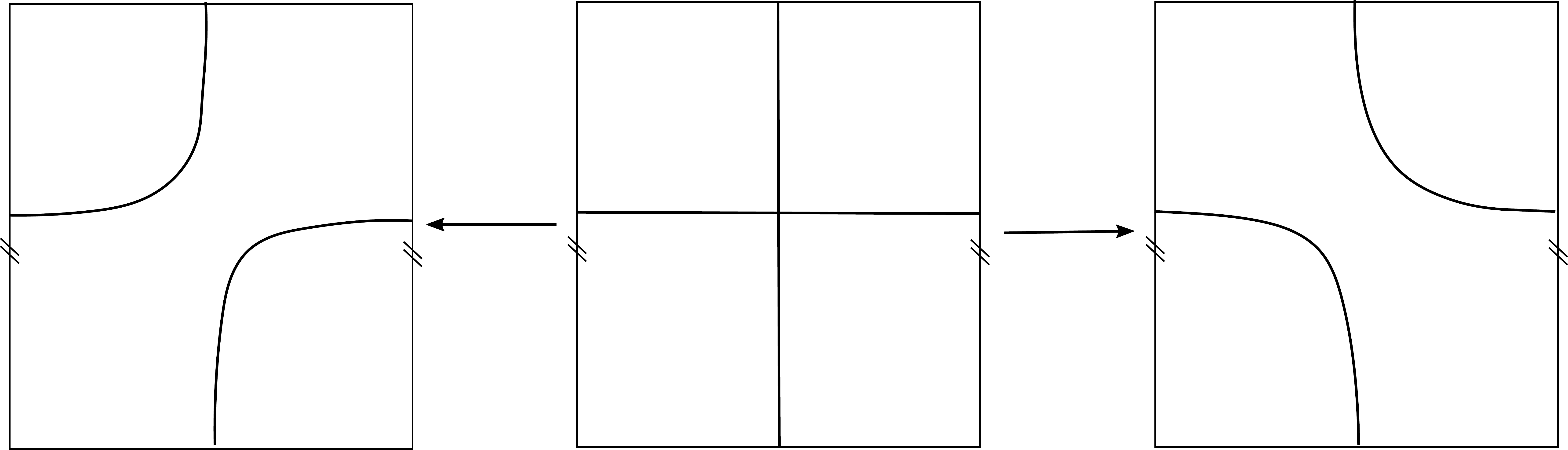}
\caption{When the $\mathbb{Z}_2$ is anomalous (i.e.\ $\alpha=-1$), the two resolutions (sides) of an intersection of lines (center) lead to different configurations.  
}\label{fig:Lpm}
\end{figure}

The phase $\alpha$ is the 't Hooft anomaly of the $\mathbb{Z}_2$ global symmetry, which is classified by the group cohomology $H^3(\mathbb{Z}_2,U(1)) = \mathbb{Z}_2$.   When $\alpha=+1$, the $\bZ_2$ is non-anomalous, while when $\alpha=-1$, the $\bZ_2$ is anomalous. One way to see this is to consider the torus partition function of the would-be $\bZ_2$ orbifold theory. The orbifold partition function involves four terms with different $\bZ_2$ line configurations, one of which is depicted in the center of Figure \ref{fig:Lpm}.  This configuration is generally ambiguous because of the intersection of the symmetry lines.  The two possible resolutions of the intersection (shown at the two sides of Figure \ref{fig:Lpm}) give the same answer if $\alpha=+1$, but not otherwise. Hence, the orbifold partition function is well-defined if $\alpha=+1$, but ambiguous and not modular invariant if $\alpha=-1$.   In the latter case $\alpha=-1$, there is therefore an obstruction to gauging the $\bZ_2$.

Another important consequence of the 't Hooft anomaly is a selection rule on the spin $s=h-\bar h$ of the state in the defect Hilbert space. For a $\bZ_2$ line, the spin of a state in ${\cal H}_{\cal L}$ is constrained to be \cite{Chang:2018iay,Lin:2019kpn}  (see also \cite{Hung:2013cda})
\ie\label{spinselection}
s\in\begin{cases}
{\bZ\over 2}\,,~~~~~~~\,~~~~\text{if}~~\alpha=+1~(\text{non-anomalous})\,,\\
\frac 14+{\bZ\over 2}\,,~~~~~~\text{if}~~\alpha=-1~(\text{anomalous})\,.
\end{cases}
\fe
To argue for this result, we consider a $2\pi$ rotation of a state in $\mathcal{H}_{\mathcal{L}}$ viewed in the operator picture of Figure \ref{fig:HL}.  By definition, this encodes the spin of the operator.   
\begin{equation}
\includegraphics[width=.6\textwidth]{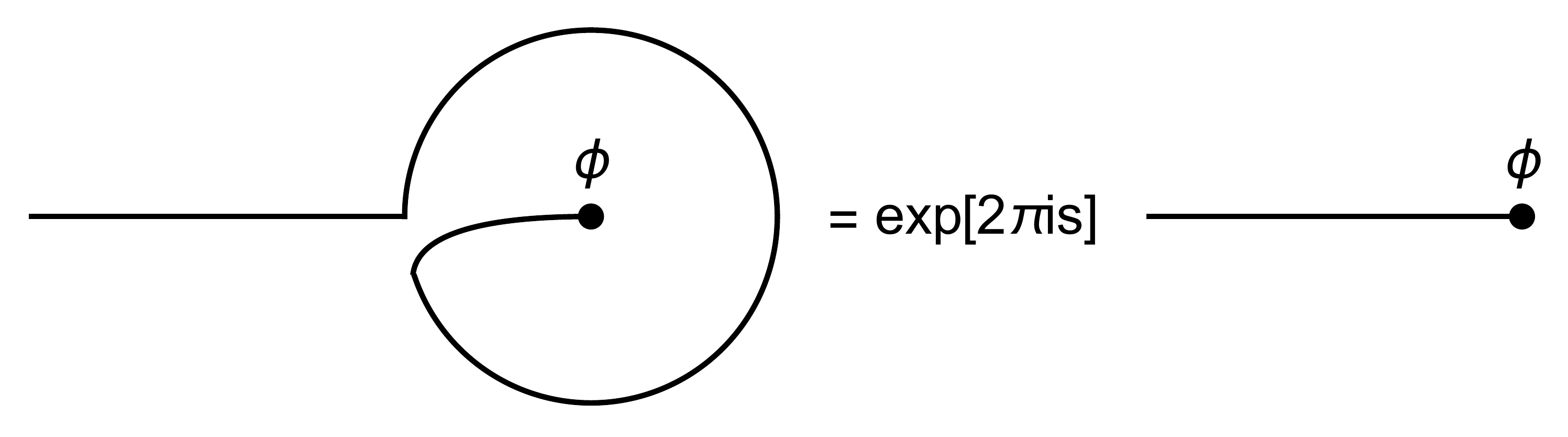}
\end{equation}
By performing this operation twice and using the crossing relation of Figure \ref{fig:crossing} we deduce that for all states in $\mathcal{H}_{\mathcal{L}},$ the spin obeys $\exp(4\pi i s)=\alpha,$ thus reproducing \eqref{spinselection}.

We can also see that the anomaly controls the expectation value of $\langle \mathcal{L} \rangle _{\mathbb{R}^{2}}$ of a closed loop of symmetry line in flat space.  Indeed, using the crossing relation of Figure \ref{fig:crossing}, we obtain the following
\begin{equation}\label{r2vev}
\includegraphics[width=.4\textwidth]{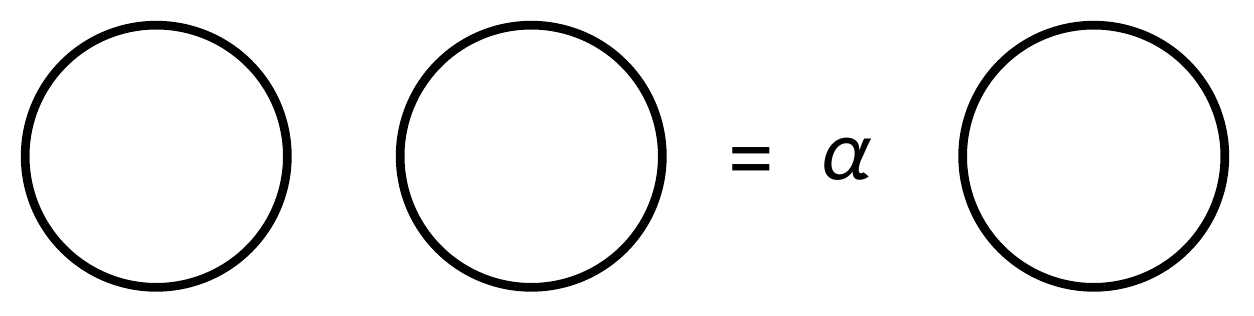}
\end{equation}
And therefore $\langle \mathcal{L} \rangle _{\mathbb{R}^{2}}=\alpha.$  We describe this equation in more detail in Section \ref{orientanom} below.

\subsection{Orientation-Reversal Anomaly on the Symmetry Line}

We now discuss time-reversal symmetry on the line $\mathcal{L}$.  As we will show, the  $\mathcal{L}$ has a $\mathsf{T}$ anomaly if and only if the bulk has the $\mathbb{Z}_{2}$ anomaly.  As a preliminary, we must first discuss why the line $\mathcal{L}$ has $\mathsf{T}$ symmetry to begin with.  Since $\mathsf{T}$ symmetry acts geometrically by reversing orientation this means we must discuss the orientation of the symmetry lines.  

In general when defining correlation functions in QFTs with global symmetry, the associated symmetry defects must be oriented.  Let $\mathcal{L}$ denote such an oriented defect, and $\overline{\mathcal{L}}$ the defect with opposite orientation.  The meaning of the choice of orientation is that if $\mathcal{L}$ is associated to the symmetry group element $g$ then $\overline{\mathcal{L}}$ is associated to the symmetry group element $g^{-1}$.  From this point of view, the reason why $\mathbb{Z}_{2}$ symmetry is special is that the non-trivial element is its own inverse.  Therefore, up to anomalies discussed below, there is no dependence on the orientation of $\mathcal{L}$; i.e.\ for $\mathbb{Z}_{2}$ lines $\mathcal{L}\cong \overline{\mathcal{L}}$.  This in essence is what defines the $\mathsf{T}$ symmetry of the defect $\mathcal{L}$.

Intuitively, we can then view the result  \eqref{r2vev} that $\langle \mathcal{L} \rangle _{\mathbb{R}^{2}}=\alpha$ as a manifestation of a $\mathsf{T}$ anomaly on $\mathcal{L}$.  Indeed, orienting $\mathcal{L}$ and stretching it to be long and thin, we can view this result as stating that even for $\mathbb{Z}_{2}$, there can be phase ambiguities in comparing $\mathcal{L}$ and $\overline{\mathcal{L}}$.    Our goal is now to make these remarks more precise.

\subsubsection{Symmetry Lines on Curved Surfaces}

The most transparent way to define and detect a $\mathsf{T}$ anomaly on the symmetry line $\mathcal{L}$ is to consider its properties on curved surfaces.  We discuss these ideas following \cite{Chang:2018iay}.  For now we let $\mathcal{L}$ be a symmetry line for a general symmetry group and subsequently specialize to the case of $\mathbb{Z}_{2}$.

As emphasized in the introduction, on a \textit{flat} manifold, the correlation functions of symmetry lines  are invariant under any deformation of the line ${\cal L}$ as long as it does not pass through any local operators.  What about on a curved  orientable surface?  In general in this situation, we should expect that the topological nature of the symmetry lines may be modified by phases.  This is because there may be a contact term in the OPE between the energy-momentum tensor and the topological line
\begin{equation}\label{lineope}
T(x,y) \mathcal{L}\sim \theta_{\cal L} \delta'(y)~, 
\end{equation}
where the line is placed at $y=0$, and $\theta_{\mathcal{L}}$ is a constant that depends on the line.  In the presence of a general metric, which acts as a source for $T$ this leads to phase modifications when $\mathcal{L}$ is deformed.  

\begin{figure}[h!]
	\centering
	\includegraphics[width=.4\textwidth]{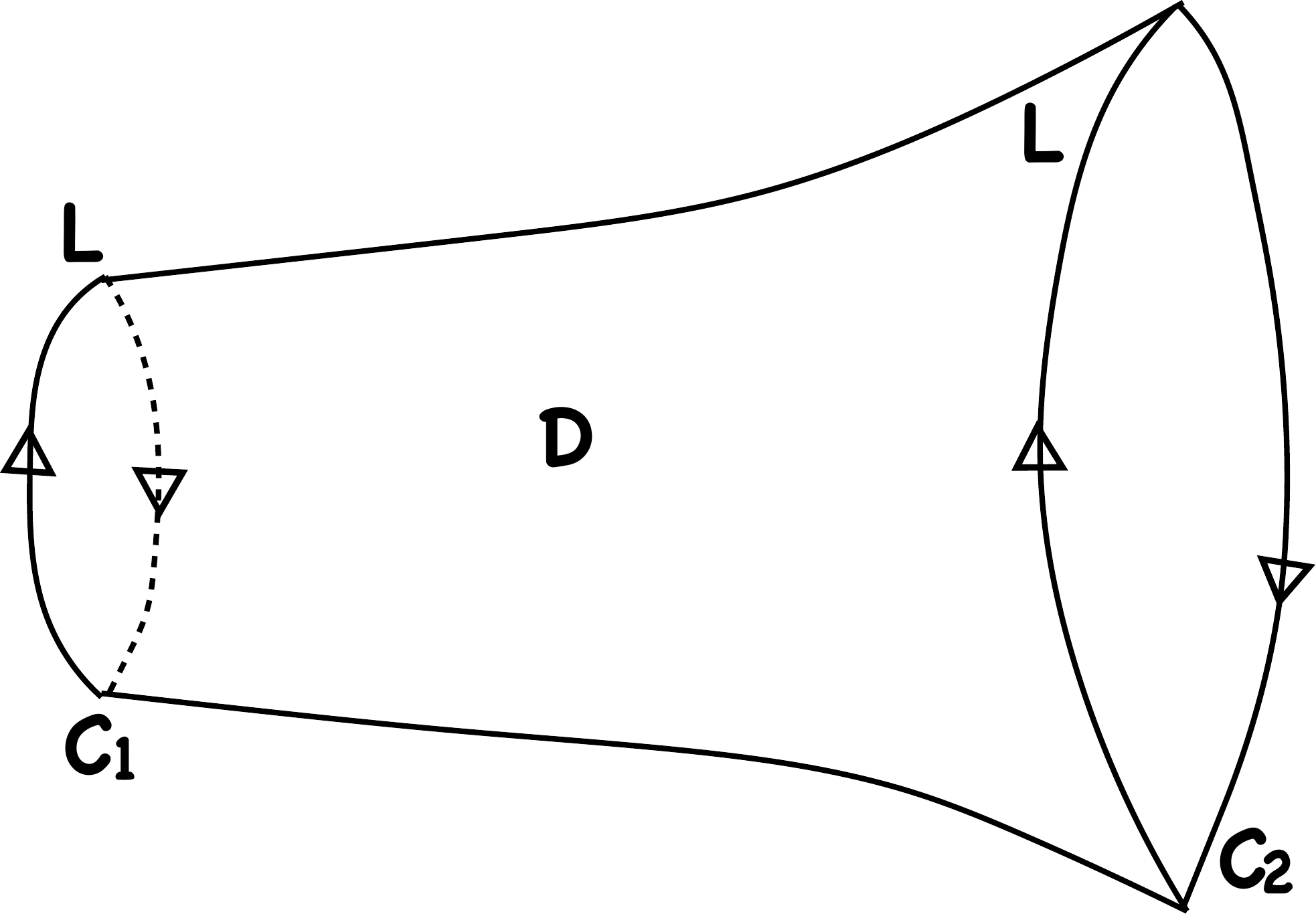}
	\caption{The line $\mathcal{L}$, initially supported on an oriented curve $C_1$, is deformed past a domain $D$ to reach the final locus $C_2$.}
	\label{fig:phasemodification}
\end{figure}

To see the consequences of \eqref{lineope}, let the initial locus of the line $\cal L$ be an oriented curve $C_1$, and deform 
$\cal L$ past a domain $D$ to reach the final locus, another oriented curve $C_2$. This is illustrated in Figure \ref{fig:phasemodification}.

The correlation functions are then modified by a phase\footnote{ We fix conventions so that orientation of $D$ is the same as the ambient surface.  The boundary of the domain $D$ is then either  $\partial D = C_1 \cup \bar C_2$ or $\partial D =\bar  C_1 \cup  C_2$, depending on which direction we move the line on the surface.}  
\begin{align}\label{phase}
\exp\left[\pm {i \theta_{\cal L} \over 4\pi} \int _D d^2\sigma \sqrt{g}R\right]\,,
\end{align}
where the sign above is positive (negative) if $\partial D$ includes $C_{1}$ with positive (negative) orientation, and $R$ is the Ricci scalar curvature on the surface.\footnote{Our normalization is such that the Gauss-Bonnet formula reads $\int_{\Sigma} d^2\sigma \sqrt{g}R(g)=4\pi \chi(\Sigma)$ with $\chi$ the Euler characteristic. }   Alternatively we can also evaluate the effects of the same deformation using the orientation reversed line $\overline {\cal L}$  starting from  $\bar C_1$ (the curve $C_1$ with opposite orientation), and ending at $\bar C_2$.  Demanding that we arrive at the same phase implies that
\begin{equation}
\theta_{\bar{\cal L}} = - \theta_{\cal L}~.
\end{equation}
 
The variation of the correlation function by the phase \eqref{phase} has the appearance of an anomaly, i.e. a violation of the topological nature of the lines by a phase.  However to check whether it is in fact meaningful, we must now determine if we can modify the definition of the symmetry line $\mathcal{L}$ by a suitable counterterm to remove this phase.  

Indeed, the phase  \eqref{phase} is not physical and can be removed using a counterterm precisely when ${\cal L}\neq \overline {\cal L}$.  Specifically, we can redefine the line $\cal L$ by including
\begin{align}\label{counter}
\exp\left[   {i\phi_{\cal L} \over 2\pi} \int_{\cal L} ds K\right]\,,
\end{align}
where $K$ is the extrinsic  curvature, normalized such that the counter-clockwise integral along the boundary of a disc is $2\pi$, and for consistency we define $\phi_{\overline{\cal L}} = -\phi_{\cal L}$. Applying the Gauss-Bonnet theorem to the cylindrical region $D$ that occurs in the calculation of \eqref{phase} we have
 \begin{align}
{1\over 4\pi}   \int _D d^2\sigma \sqrt{g}R(g) +{1\over 2\pi} \int _{\partial D} ds K = \chi(D)=0~,
  \end{align}
where the last equality follows from the fact that $D$ is an annulus.  Thus, the phase \eqref{phase} is canceled by the counterterm \eqref{counter} if we choose $\phi_{\cal L} =  \theta_{\cal L}$.

In particular, we note that while the phase \eqref{phase} is only nontrivial on a curved manifold, the counterterm \eqref{counter} is generally nonzero even on $\mathbb{R}^2$.  (In this case the extrinsic curvature $K$ simply integrates to the total angle $2\pi$ around any closed curve).  Therefore the counterterm \eqref{counter} changes the expectation value of an closed loop $\langle {\cal L}\rangle_{\mathbb{R}^2}$ by the arbitrary phase $\exp(i \phi_{\cal L}).$  Thus, when ${\cal L}\neq \overline {\cal L}$ the expectation value $\langle {\cal L}\rangle_{\mathbb{R}^2}$ does not have intrinsic meaning.\footnote{Mathematically, one can think of this extrinsic curvature counterterm as representing some of the exact elements that are set to zero in defining the group cohomology $H^3(G,U(1))$.  Specifically, it represents those two-cochains of the form $\beta(g,g^{-1})/\beta(g^{-1},g)$ for $g$ in the zero-form group $G$.}

\subsubsection{Orientation-Reversal Anomaly for $\mathbb{Z}_{2}$ Lines}\label{orientanom}

Let us now turn to the special case of most interest, which is $\mathbb{Z}_{2}$ global symmetry.  As we have discussed above, in this case locally, the symmetry line $\mathcal{L}$ is invariant under changes in orientation.  This statement should be interpreted as an equality between $\mathcal{L}$ and $\overline{\mathcal{L}}$ on $\mathbb{R}^{2}$.  In particular, this requirement forbids us from modifying the line by the extrinsic curvature counterterm \eqref{counter}.  Thus, precisely for $\mathbb{Z}_{2}$ lines the expectation value $\langle {\cal L}\rangle_{\mathbb{R}^2}$ discussed in \eqref{r2vev} is meaningful. 

Since the counterterm \eqref{counter} is unavailable, the phase \eqref{phase} is now an intrinsic physical property of the line.  To interpret its effect we reexamine Figure \ref{fig:phasemodification}.  The boundary of the cylindrical region $D$ is two isotopic lines with opposite orientation.  Therefore, we can view the phase \eqref{phase} as the cost of changing the orientation of the symmetry line $\mathcal{L}$ on a curved manifold.  In particular, a non-zero phase $\theta_{\mathcal{L}}$ implies that on curved manifolds, in contrast to $\mathbb{R}^{2}$, the correlation functions of $\mathcal{L}$ depend on its orientation.  This is an orientation-reversal or $\mathsf{T}$ anomaly on the defect $\mathcal{L}$.  
 
We now show that this orientation reversal anomaly is precisely controlled by the bulk $\mathbb{Z}_{2}$ anomaly thus demonstrating the correspondence \eqref{bossmith} in this case.  First, note that the phase $\theta_{\cal L}$ is constrained by \eqref{r2vev}.  Consider a small loop of $\cal L$ on a two-sphere.  Near the north pole, the expectation value is approximated by  $\langle {\cal L}\rangle_{\mathbb{R}^2}$, while near the south pole it is $\langle \overline {\cal L}\rangle_{\mathbb{R}^2}$.  Deforming between them using the formula \eqref{phase} we find
\begin{align}
\langle {\cal L}\rangle_{\mathbb{R}^2}  = \exp(\pm 2 i \theta_{\cal L})\,\langle \overline {\cal L}\rangle_{\mathbb{R}^2} \,.
\end{align}
Therefore, since $\langle {\cal L}\rangle_{\mathbb{R}^2}=\langle \overline {\cal L}\rangle_{\mathbb{R}^2}$ we deduce that $\theta_{\mathcal{L}}$ is either zero or $\pi$.  In the former case there is no $\mathsf{T}$ anomaly on the line, while in the latter case there is a $\mathsf{T}$ anomaly.

\begin{figure}
\centering
\includegraphics[width=.3\textwidth]{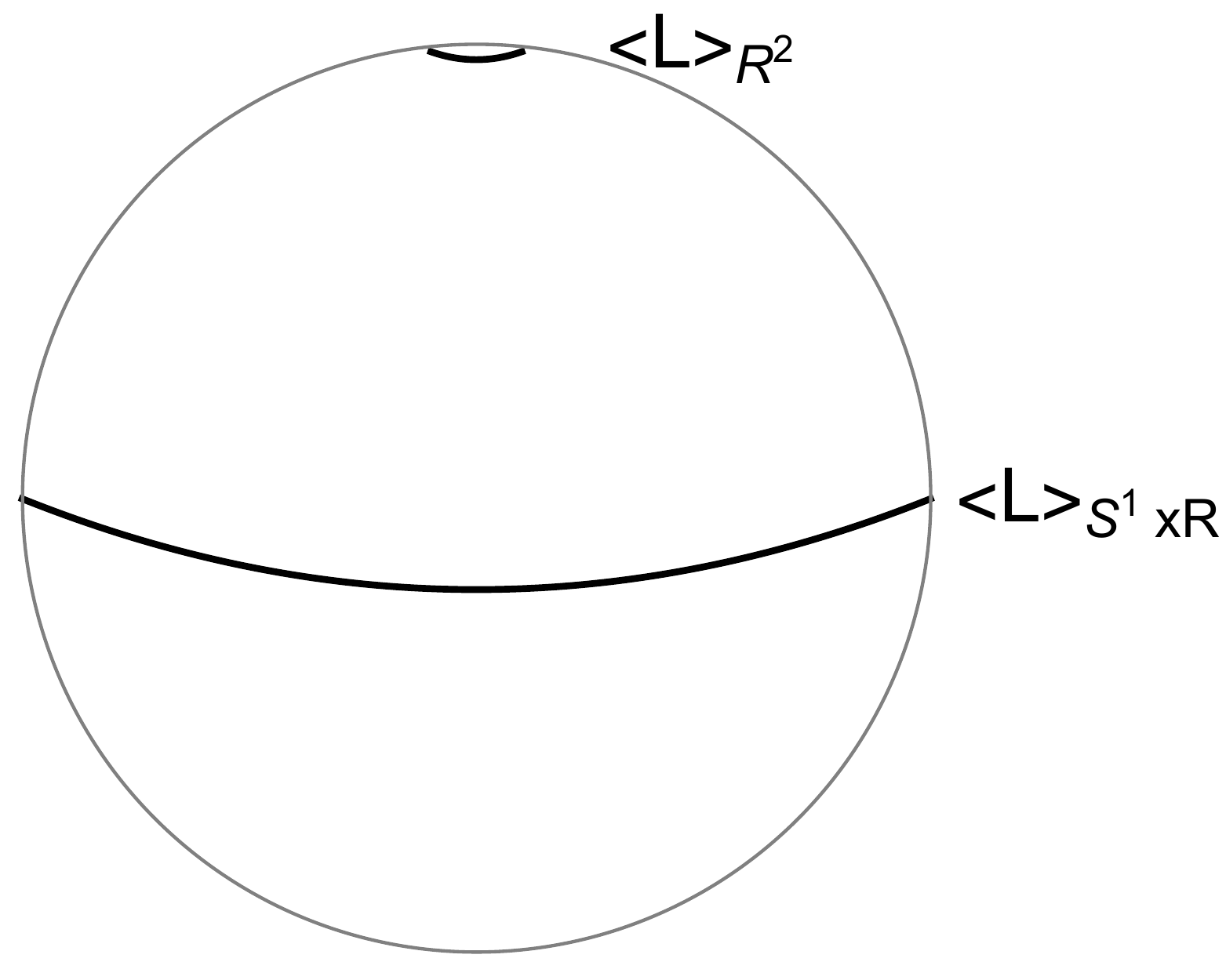}
\caption{The expectation value of an anomalous $\bZ_2$ changes as we deform it on a curved manifold. For example, the vev near the north pole of a two-sphere differs from that around the equator by a sign.}\label{fig:sphere}
\end{figure}

Similarly, we can compare $\mathcal{L}$ wrapping the equator of the two-sphere to its behavior near a pole.  On the equator, the expectation value looks locally like $\mathcal{L}$ wrapping the non-trivial cycle on the cylinder $S^{1}\times \mathbb{R}$.  
This expectation value is easy to determine.  Since there is no operator insertion, we can view the defect $\mathcal{L}$ as acting on the vacuum in Hilbert space $\mathcal{H}$ of states on $S^{1}$. In any unitary theory, the vacuum on the circle is unique and symmetry preserving and therefore $\langle{ \cal L}\rangle_{S^1\times \mathbb{R}} = +1.$  Therefore, deforming between the equator and a pole (see Figure \ref{fig:sphere}) and using the formula \eqref{phase} we derive:
\begin{align}
\langle {\cal L}\rangle_{ \mathbb{R}^2} =
 \exp(i \theta_{\cal L}) \, 
\langle {\cal L}\rangle_{S^1\times \mathbb{R}}  \,.
\end{align}
Combining \eqref{r2vev} and the above, we conclude that: \textit{In a unitary theory,  the $\mathbb{Z}_2$ line $\cal L$ has the  orientation-reversal} ($\mathsf{T}$) \textit{anomaly if and only if the $\mathbb{Z}_2$ is anomalous}:\footnote{This derivation requires unitarity.  Indeed, in a non-unitary theory, the cylinder expectation value of a symmetry line may not be +1.  Then the above relation is modified to  $ \exp(i \theta_{\cal L}) =  {\alpha\over 
\langle {\cal L}\rangle_{S^1\times \mathbb{R}}  }$. 
For example, if $\langle {\cal L}\rangle_{S^1\times \mathbb{R}}=-1$ but $\alpha=-1$, then the anomalous $\mathbb{Z}_2$ line does not have the orientation-reversal  anomaly in this non-unitary theory.  Such an anomalous $\mathbb{Z}_2$ without an orientation-reversal anomaly on the symmetry line can be found in, for example, the (3,5) minimal model.}
\begin{align}
\exp({ i\theta_{\cal L } }  )=  \alpha\,.
\end{align}

\section{Kramers Doublets in the Defect Hilbert Space}

We now illustrate the conclusion of the previous section in several examples.  As discussed above, we interpret the orientation-reversal anomaly of an anomalous $\mathbb{Z}_2$ line  as the time-reversal anomaly for the defect Hilbert space ${\cal H}_{\cal L}$.  This gives a direct physical realization of the isomorphism \eqref{bossmith} in the special case of $(1+1)d$ bosonic $\bZ_2$ anomalies using the language of topological defect lines. 

A $(0+1)d$ bosonic theory with $\mathsf{T}$ symmetry has an anomaly exactly when $\mathsf{T}$ acts projectively, i.e. $\mathsf{T}^2=-1$ on ${\cal H}_{\cal L}$.  This results in the Kramers doublet degeneracy in the defect Hilbert space ${\cal H}_{\cal L}$.  Below we will demonstrate in examples that the defect Hilbert space ${\cal H}_{\cal L}$ has the Kramers degeneracy if and only if the bulk $\bZ_2$ symmetry is anomalous. Most of our examples are diagonal rational CFTs, whose  defect Hilbert space ${\cal H}_{\cal L}$ degeneracies are given by the fusion coefficients $N^i_{jk}$ \cite{Petkova:2000ip} (see also \cite{Chang:2018iay}).

\subsection{$\widehat{\mathfrak{su}(2)}_1$ WZW Model}

The $\widehat{\mathfrak{su}(2)}_1$ WZW model has two current algebra primaries, the vacuum $|0,0\rangle$ and the four spin one-half primaries $|\frac 14 , \frac14\rangle_{\pm,\pm}$ of weight $h=\bar h =\frac 14$.  
The $\pm$ stands for the $J^3_0$ and $\bar J^3_0$ eigenvalues. 
This theory has an anomalous $\mathbb{Z}_2$ global symmetry ($\alpha=-1$) which commutes with the $\widehat{\mathfrak{su}(2)}\times \widehat{\mathfrak{su}(2)}$  current algebra, and acts on the primaries as
\begin{align}
\mathbb{Z}_2:~~|0,0\rangle  \mapsto |0,0\rangle\,,~~~~|\frac14,\frac14\rangle_{\pm,\pm} \mapsto -| \frac14,\frac14\rangle_{\pm,\pm}\,.
\end{align}
This $\mathbb{Z}_2$ can be thought of as the center of the left $\widehat{\mathfrak{su}(2)}$.

The defect Hilbert space ${\cal H}_{\cal L}$ of this anomalous $\bZ_2$ has been studied, for example, in \cite{Lin:2019kpn}.  It consists of two current algebra primaries with $h=\frac 14 $ and $\bar h=0$, and another two with $h=0$ and $\bar h=\frac14$.  Each one of them is a doublet under either the left or the right $\widehat{\mathfrak{su}(2)}$ current algebra:
\ie
{\cal H}_{\cal L}:~~|  \frac 14 , 0\rangle_\pm\,,~~~~|  0,\frac 14 \rangle_\pm\,.
\fe
Note that they have fractional spin $s=h-\bar h = \pm \frac14$, consistent with the spin selection rule for states in ${\cal H}_{\cal L}$  \eqref{spinselection}.  Morevoer, we indeed see that there is a two-fold degeneracy for every state of a given $(h,\bar h)$ which is the expected Kramers degeneracy.

\subsection{$T^2$ CFT with no Time-Reversal Symmetry}

In the general construction in Section \ref{bosonic}, the $(1+1)d$ theory is assumed to have a $\bZ_2$ global symmetry, but is not in general time-reversal invariant. The $\widehat{\mathfrak{su}(2)}_1$ WZW model has time-reversal symmetry, and here we discuss an example without.\footnote{Another  example is  the tensor product of the $\widehat{\mathfrak{su}(2)}_1$ WZW model with the holomorphic Monster CFT.  This theory does not have $(1+1)d$ time-reversal symmetry and it has an anomalous $\mathbb{Z}_2$ in the $\widehat{\mathfrak{su}(2)}_1$ WZW sector.  Its defect Hilbert space is what we described in the previous section tensor product with the Monster CFT, which is not invariant under $h\leftrightarrow \bar h$ but still has the Kramers degeneracy in states with the same $h$ and $\bar h$.}

The example we will consider is  the free CFT with target space being a two-dimensional torus $T^2$, with a general metric and $B$ field. We will follow the convention in \cite{Polchinski:1998rq} with $\alpha'=1$. 
 The two scalar fields of the $T^2$ CFT are normalized to have periodicities
\begin{align}
X^1 \sim X^1+2\pi R\,,~~~~X^2 \sim X^2+2\pi R\,.
\end{align}
The metric and the $B$ field of the $T^2$ CFT will be denoted as $G_{mn}$ and $B_{mn}$ with $m,n=1,2$, parametrizing the moduli space of the $T^2$ CFT.   Since we only have two scalars, there is only one $B$ field, $B\equiv B_{12}$.  

The metric moduli includes the overall volume and the complex structure moduli $\tau$. The overall volume can be parametrized by the $R$ above, while the complex structure $\tau$ is encoded in $G_{mn}$ as
\begin{align}
G_{mn} = \left(\begin{array}{cc}1 & \tau_1 \\\tau_1 & |\tau|^2\end{array}\right)\,,
\end{align}
where $\tau = \tau_1 + i\tau_2$ and $|\tau|^2 = \tau_1^2 +\tau_2^2$.

The spectrum of local operators can be described as follows. Let
\begin{align}
v_m \equiv { n_m \over R} - B_{mn} w^n R\,,
\end{align}
where $n_m,w^n\in \mathbb{Z}$ are the momentum and winding numbers.   Next we define
\begin{align}
v_{L}^m  = v^m + w^m R\,,~~~~~~v_R^m  = v^m -w^mR\,,
\end{align}
where the indices are raised and lowered by $G_{mn}$ and $G^{mn}$. For example, $v^m = G^{mn} v_n$.  
The current algebra primaries in the $T^2$ CFT are labeled by four integers, $(n_1,n_2,w^1,w^2)$, with weights
\begin{align}
h  = \frac14 G_{mn} v^m_L v^n_L\,,~~~~~\bar h  = \frac 14 G_{mn} v^m_Rv^n_R\,.
\end{align}
At a generic point on the moduli space with nontrivial $B$ field, the $(1+1)d$ theory does not have a time-reversal symmetry which exchanges $h$ and $\bar{h}$.

There are various anomalous $\mathbb{Z}_2$ symmetries.  We will focus on the anomalous $\mathbb{Z}_2$ subgroup of $\text{diag}(U(1)_{n_1}\times U(1)_{w^1})$.  
The twisted sector operators are those with $n_1,w^1\in \frac 12+\bZ$ and $n_2,w^2\in \bZ$ (see, for example, \cite{Lin:2019kpn,Ji:2019ugf}). We therefore find that the defect Hilbert space has a two-fold Kramers degeneracy in states with the same $h$ and $\bar h$, which is realized by the action on the mode numbers
\begin{align}
(n_1,n_2,w^1,w^2)  \to (-n_1-1 ,-n_2 , -w^1-1 , -w^2)\,.
\end{align}

\subsection{Ising Model}\label{Sec:IsingExample}

Finally, let us consider an example where the $(1+1)d$ $\bZ_2$ symmetry is non-anomalous, and show that the defect Hilbert space has no Kramers degeneracy. 

The $(1+1)d$ Ising model has  three Virasoro primaries, the vacuum $1$ with $h=\bar h=0$, the energy operator $\epsilon$ with $h=\bar h=\frac 12$, and the spin field $\sigma$ with $h=\bar h={1\over 16}$.  There is a  $\bZ_2$ symmetry   that flips the sign of the spin field
\begin{align}
\bZ_2:~~1\to 1\,,~~~~\epsilon\to\epsilon\,,~~~~\sigma\to -\sigma\,.
\end{align}
The $\bZ_2$ symmetry is non-anomalous $\alpha=+1$. Indeed, the Ising CFT is self-dual under the $\bZ_2$ orbifold.

The defect Hilbert space ${\cal H}_{\cal L}$ of the $\bZ_2$ line consists of following Virasoro primaries $|h,\bar h\rangle$ (see, for example, \cite{Lin:2019kpn}):
\begin{align}
\bar\psi(\bar z) = | 0,\frac12\rangle, ~~~\psi(z) = |\frac 12,0\rangle,~~ ~\mu(z,\bar z)=|\frac {1}{16},{1\over 16}\rangle\,.
\end{align}
Under the operator-state correspondence, they are mapped to operators living at the end of the $\bZ_2$ line. 
The $|0,\frac 12\rangle$ and $|\frac 12,0\rangle$ states are the free Majorana fermions $\psi(z), \bar\psi(\bar z)$ with half integral spins. 
The $|{1\over 16},{1\over 16}\rangle$ state in ${\cal H}_{\cal L}$ is the disorder operator $\mu(z,\bar z)$, which is not mutually local with the spin field $\sigma(z,\bar z)$ in $\cal H$ because the latter is $\bZ_2$ odd.  
Note that the spins of these states are consistent with the spin selection rule \eqref{spinselection}. 
We see that there is no Kramers degeneracy for the states in ${\cal H}_{\cal L}$, hence there is no $\mathsf{T}$ anomaly in this case.

\section{Fermionic Examples in $(1+1)d$}\label{fermsec}

In this section we give an example illustrating the correspondence \eqref{bossmith} for spin theories in $(1+1)d$.  As we review below, in this case possible anomalies for $\mathbb{Z}_{2}$ are classified by $\mathbb{Z}_{8}$ \cite{PhysRevB.81.134509,PhysRevB.85.245132,Qi_2013,PhysRevB.88.064507,PhysRevB.89.201113, Kapustin:2014dxa}.  Similarly, in $(0+1)d,$ $\mathsf{T}$ anomalies for theories with $\mathsf{T}^{2}=1$ (i.e.\ $\text{pin}^{-}$ structure) are also classified by $\mathbb{Z}_{8}$ \cite{Fidkowski_2011, PhysRevB.81.134509, Kapustin:2014dxa, Freed:2016rqq}.\footnote{Mathematically, the $(1+1)d$ SPT associated to the $\mathsf{T}$-invariant fermions is the $\mathbb{Z}_8$-valued Arf-Brown-Kervarie cobordism invariant of $\text{pin}^{-}$ two-manifolds.  Meanwhile the $(2+1)d$ $\mathbb{Z}_{2}$ SPT can be constructed by evaluating an $SO(N)_{1}$ Chern-Simons path-integral coupled to a background magnetic field \cite{Cordova:2017vab}.  }  The correspondence \eqref{bossmith} then directly relates these results.  We concentrate our analysis on free fermion systems and their interactions.  

\subsection{$\mathbb{Z}_{2}$ Invariant $(1+1)d$ Fermions}

Consider a $(1+1)d$ system of $N$ free (non-chiral) Majorana fermions $\chi^{i}(z,\overline{z})$.  
We will denote their left- and right-moving components as $\psi^{i}(z), \rho^{i}(\bar{z})$, respectively.  We define the global $\mathbb{Z}_{2}$ symmetry of interest to be the left-moving fermion number $(-1)^{F_{L}}$ which acts on $\psi^{i}$, but under which the right moving fermions $\rho^{i}(\bar{z})$ are neutral.  There is also the total fermion number symmetry that acts diagonally on both left and right-moving fields.  These $N$ fermions have an $SO(N)\times SO(N)$ global symmetry under which $\psi$ and $\rho$ transform respectively as $(v,1)$ and $(1,v)$ where $v$ is the vector representation.  Under the diagonal $SO(N)\subset SO(N)\times SO(N)$ the Majoranas $\chi^{i}$ are thus also in the vector representation.

The $(-1)^{F_{L}}$ symmetry prohibits quadratic mass terms $\psi^{i}(z)\rho^{j}(\bar{z})$ and hence, at the quadratic level, the number of fermions $N$ is protected.  We now investigate to what extent this statement survives interactions.  Consider a quartic interaction invariant under the diagonal $SO(N)$:
\begin{equation}\label{GNop}
H_{int}= q \left(\sum_{i=1}^{N}\bar{\chi}^{i}\chi^{i}\right)^{2}~.
\end{equation}
Following \cite{PhysRevB.85.245132,Qi_2013,PhysRevB.88.064507,PhysRevB.89.201113}, our goal is to see to when such interactions can lead to a trivially gapped state preserving the chiral symmetry.  Our treatment closely parallels the review  \cite{Tong:2019bbk}.

For two fermions, the first case where this interaction is possible, the free theory describes a $c=1$ model and the quartic fermion interaction is the unique exactly marginal operator.  Therefore in this case the theory remains gapless for all $q$.  For $N>2$, the operator \eqref{GNop} is marginal but not exactly marginal.  For one sign of $q$ this leads again to the gapless free fermions, however for the other sign of $q$, the interaction operator becomes important at long distances leading to the $SO(N)$ Gross-Neveu model \cite{Gross:1974jv}.  This model is gapped and at a strong coupling scale $\Lambda$, generates an expectation value 
\begin{equation}\label{GNvev}
\langle\sum_{i=1}^{N}\bar{\chi}^{i}\chi^{i}\rangle=\pm \Lambda~.
\end{equation}
This expectation value spontaneously breaks the $(-1)^{F_{L}}$ symmetry leading to two degenerate ground states.

Something special happens with eight fermions however, that will enable us to find an appropriate interaction to gap the system.  The special feature that we need is triality \cite{Witten:1984mb}. This allows us to present the theory of the eight fermions in dual description where there are again eight Majorana fermions $\tilde{\chi}^{m}$ ($m=1,\cdots,8$) but now the left- and right-moving parts transform under $SO(8)\times SO(8)$ as $(s,1)$ and $(1,s)$ where $s$ is an eight-dimensional spinor representation.  These spinor fermions are coupled to a dynamical $\mathbb{Z}_{2}\times \mathbb{Z}_{2}$ gauge theory with gauge fields $x$ and $y$ coupling respectively to the $(-1)^{\tilde{F}}$ and $(-1)^{\tilde{F}_{L}}$ symmetries of the fermions $\tilde{\chi}^{m}$.  Triality has an important interplay with operators in various sectors of the theory. In the $\chi$ theory, there are spin fields $\sigma^{m,n}$ which are $(R,R)$ sector operators of dimension $(1/2,1/2)$ and transform as $(s,s)$ under the $SO(8)\times SO(8)$ global symmetry.  Crucially for our purposes, under triality these spin fields are exchanged with bilinears in the $\tilde{\chi}$ fermions.\footnote{In fact the existence of the triality shows that the chiral symmetry  $(-1)^{\tilde F_L}$ of the $\tilde \chi$ theory, itself a theory of eight fermions, is non-anomalous in the sense that it can be gauged. Here we simply show how to use this triality to trivially gap the original model. To derive this triality from first principles, one can start with eight fermions in the vector representation of $SO(8)$, and sum over the spin structures (i.e. bosonization) to obtain the (bosonic) $Spin(8)_1$ WZW model. The latter has a non-anomalous $\bZ_2\times \bZ_2$ global symmetry (see Appendix B of \cite{Lin:2019kpn}).  Next, the fermionizations  (in the sense of \cite{Gaiotto:2015zta,Kapustin:2017jrc,Thorngren:2018bhj,Karch:2019lnn,Ji:2019ugf}) of the $Spin(8)_1$ WZW model with respect to the three order two elements give eight fermions in the vector, spinor, conjugate spinor representations of the $SO(8)$.  Therefore, the triality can be realized as a composition of bosonization and fermionization.}  

Since we are interested in gapping the model while preserving the original $(-1)^{F_{L}}$ symmetry, we should track how this behaves under the above triality.  Let $a$ be the background gauge field for the $(-1)^{F_{L}}$. In terms of the original vector-valued fermion variables $\chi$, the coupling to $a$ appears in the kinetic term.  However, in the dual theory with $SO(8)$ spinor fermions $\tilde{\chi}$, the symmetry $(-1)^{F_{L}}$ does not couple directly to the fermions but rather couples to the twisted sector operators that emerge from the dynamical $\mathbb{Z}_{2}\times \mathbb{Z}_{2}$ gauge fields $x$ and $y$.  Specifically this means that $a$ enters the action for the dual theory as $\exp(i\pi a\cup(x+y))$, where $x,y,a \in H^{1}(\Sigma,\mathbb{Z}_{2})$ and $\Sigma$ is spacetime.

We now return to our original goal of trivially gapping the theory of eight fermions $\chi$.  An enlightening attempt, which ultimately will not suceed, is to try to add classically marginal potential which is invariant under the diagonal $SO(8)\subset SO(8)\times SO(8)$ constructed out of a quadratic combination spin fields $\sigma^{m,n}$  transforming in the $(s,s)$. Using the triality discussed above, we can understand the effects of this interaction by mapping it to a quartic fermion operator in the dual fields $\tilde{\chi}$:
\begin{equation}\label{spintry}
H_{int} =q \left(\sum_{m=1}^{8}\sigma^{m,m}\right)^{2}\longleftrightarrow \tilde{H}_{int}=q \left(\sum_{m=1}^{8}\bar{\tilde{\chi}}^{m}\tilde{\chi}^{m}\right)^{2}~.
\end{equation}
From the right-hand-side, it is clear that for one sign of $q$ we again flow to an $SO(8)$ Gross-Neveu model.  Since the broken chiral symmetry $(-1)^{\tilde{F}_{L}}$ is gauged in the $\tilde{\chi}$ theory, this leads to a unique ground state in this sector.  However, there remains the unbroken $(-1)^{\tilde{F}}$ symmetry which is also gauged and survives at long distances.  Moreover, in the $\tilde{\chi}$ theory $a$ couples to the twisted sector of the surviving dynamical $\mathbb{Z}_{2}$ gauge field.  Thus, the interaction \eqref{spintry} leads to two grounds states and again spontaneously breaks $(-1)^{F_{L}}.$

Finally, let us try an interaction preserving only the diagonal $SO(7)$
\begin{equation}\label{spintry1}
H_{int} =q \left(\sum_{n=1}^{7}\sigma^{n,n}\right)^{2}+p\left(\sum_{n=1}^{7}\sigma^{n,n}\right)\sigma^{8,8}\longleftrightarrow \tilde{H}_{int}=q \left(\sum_{n=1}^{7}\bar{\tilde{\chi}}^{n}\tilde{\chi}^{n}\right)^{2}+p\left(\sum_{n=1}^{7}\bar{\tilde{\chi}}^{n}\tilde{\chi}^{n}\right)\bar{\tilde{\chi}}^{8}\tilde{\chi}^{8}~.
\end{equation}
Working in the limit $|q|\gg|p|$ we first flow to an $SO(7)$ Gross-Neveu model.  As above this leads to a unique ground state among these seven fermions.  What remains at low energies is therefore the final fermion $\tilde{\chi}^{8}$ coupled to a dynamical $\mathbb{Z}_{2}$ gauge field $x$ acting on $\tilde{\chi}^{8}$ as the overall fermion number $(-1)^{F}$.  Moreover, there is also a mass term arising from the chiral expectation value \eqref{GNvev} and the interaction controlled by $p$.  In summary the long-distance effective action takes the form: 
\begin{equation}
S_{eff}=\int \bar{\tilde{\chi}}^{8}D_{x}\tilde{\chi}^{8}+m \bar{\tilde{\chi}}^{8}\tilde{\chi}^{8}+i\pi x\cup a~.
\end{equation}
This is exactly the fermionic presentation of the $(1+1)d$ Ising model from  gauging  $(-1)^F$ of a non-chiral, massive Majorana fermion. 
In particular, the fermion mass operator is equivalent to the energy operator $\varepsilon$.  For one sign of the mass, the theory spontaneously breaks the $(-1)^{F_{L}}$ symmetry.  For the other sign, it is trivially gapped with a symmetric ground state as desired.\footnote{To recover the same result in the fermionic description, note that for one sign of the mass the low-energy limit of the fermions is trivial, leading to two ground states, while for the other sign of the mass, the fermion path-integral generates an $Arf(x+s)$ interaction, where $x$ is the $\mathbb{Z}_{2}$ dynamical gauge field and $s$ is the underlying spin structure. This leads to a trivially gapped theory. }

\subsection{$\mathsf{T}$ Invariant $(0+1)d$ Fermions}

Consider now a $(0+1)d$ system of $N$ free massless real fermions $\lambda^i(t).$   These fermions have a $\mathsf{T}$ symmetry that acts as
\begin{equation}
\mathsf{T} \lambda^i(t) \mathsf{T}^{-1}= - \lambda^i (-t)~ \Longrightarrow \mathsf{T}^2=+1~.
\end{equation}
We can try to add mass terms of the form $im \lambda^i\lambda^j$ with real $m$ to gap out these fermions.  However, these mass terms are forbidden by $\mathsf{T}$.   Thus, including only quadratic mass terms, the number of fermions $N$ is protected.

To go further it is helpful to investigate the Hilbert space in more detail.  For simplicity, we consider only even numbers of fermions and group them into complex pairs pairs to define fermionic creation and annihilation operators:
\begin{equation}
a_{n}=\frac{1}{\sqrt{2}}(\lambda^{2n-1}+i\lambda^{2n})~, \hspace{.5in}a_{n}^{\dagger}=\frac{1}{\sqrt{2}}(\lambda^{2n-1}-i\lambda^{2n})~,\hspace{.5in}\{a_{n}, a_{m}^{\dagger}\}=\delta_{nm}~.
\end{equation}
Each pair of fermions thus generates a Hilbert space spanned by states $|\downarrow\rangle$ and $|\uparrow\rangle$:
\begin{equation}
a | \downarrow \rangle=0~,\hspace{.25in}a^\dagger|\downarrow\rangle = | \uparrow\rangle~, \hspace{.25in}a| \uparrow \rangle= |\downarrow\rangle~, \hspace{.25in}a^\dagger|\uparrow\rangle = 0~.
\end{equation}
$\mathsf{T}$ acts on these states by $\mathsf{T}|\downarrow\rangle=|\uparrow\rangle$ and $\mathsf{T}|\uparrow\rangle=-|\downarrow\rangle.$   Thus, the free system with $N$ fermions has a ground state degeneracy of $2^{N/2}$.

Following \cite{PhysRevB.81.134509}, we now consider possible quartic fermion interactions to determine if we can find a system with a unique ground state.  With four fermions we can add a $\mathsf{T}$ invariant interaction term
\begin{equation}
H_{int}=q \lambda^{1}\lambda^{2}\lambda^{3}\lambda^{4}= -q \left(a^{\dagger}_{1}a_{1}-\frac{1}{2}\right)\left(a^{\dagger}_{2}a_{2}-\frac{1}{2}\right)~.
\end{equation}
Taking $q$ to be positive, we see that the this term energetically penalizes states where the spins are not aligned.  Therefore the fourfold ground state degeneracy is broken two a two-fold degeneracy among the states $|\downarrow\downarrow\rangle$ and $|\uparrow\uparrow\rangle$.  In particular, we still have ground state degeneracy.  Similarly, one can argue that with six fermions we cannot find interactions to fully lift the degeneracy.  

However, consider now eight fermions. We consider the interaction Hamiltonian
\begin{equation}
H_{int}=q (\lambda^{1}\lambda^{2}\lambda^{3}\lambda^{4}+\lambda^{1}\lambda^{2}\lambda^{5}\lambda^{6}+\lambda^{1}\lambda^{2}\lambda^{7}\lambda^{8}+\lambda^{3}\lambda^{4}\lambda^{5}\lambda^{6}+\lambda^{3}\lambda^{4}\lambda^{7}\lambda^{8}+\lambda^{5}\lambda^{6}\lambda^{7}\lambda^{8})-p\lambda^{1}\lambda^{3}\lambda^{5}\lambda^{7}~.
\end{equation}
We work in the limit where $q\gg p>0$.  First neglecting $p$ and reasoning as above we deduce that there is an approximate ground state degeneracy among the two states where all spins are aligned $|\downarrow\downarrow\downarrow\downarrow\rangle$ and $|\uparrow\uparrow\uparrow\uparrow\rangle.$  However now the $p$ interaction term acts to map between these two remaining states.  This splits the degeneracy leaving a unique $\mathsf{T}$ invariant ground state 
\begin{equation}
|\Omega \rangle = \frac{1}{\sqrt{2}}\left(|\downarrow\downarrow\downarrow\downarrow\rangle+|\uparrow\uparrow\uparrow\uparrow\rangle\right)~.
\end{equation}
This shows that including interactions the number of fermions is only invariant modulo eight, and hence the possible $\mathsf{T}$ anomaly is a mod eight effect.

 \subsection{Twisted Compactification}
 
 As is clear from the previous sections both $(1+1)d$ $\mathbb{Z}_{2}$ symmetric fermions and $(0+1)d$ $\mathsf{T}$ invariant fermions have a $\mathbb{Z}_{8}$ classification. But, the derivations of these two results appear somewhat different.  
The general correspondence \eqref{bossmith} provides a direct map between them by considering the theory on the $\mathbb{Z}_{2}$ symmetry line.  

Specifically, following our general discussion in Section \ref{genres}, we can access the theory of the symmetry defect by a $\mathbb{Z}_{2}$ twisted circle compactification of the free $(1+1)d$ fermions.  This twisted compactification of free fermions was also discussed in \cite{Qi_2013}.  Since the $\mathbb{Z}_{2}$ acts chirally as $(-1)^{F_L}$ in the $(1+1)d$ free fermion theory,  under this twisted compactification, the left-moving fermions $\psi^i(z)$ are periodic (R sector), while the right-moving fermions $\rho^i(\bar z)$ are anti-periodic (NS sector). In the low energy limit, we therefore obtain exactly $N$ $(0+1)d$ Majorana fermions $\lambda^i(t)$ coming from the zero modes of the $(1+1)d$ left-moving fermions.  In particular this explains the fact that the classification of these systems, including interactions, coincide.

\section{An Example in $(3+1)d$}\label{hdsec}

As a final example of the correspondence \eqref{bossmith}, we consider $(3+1)d$ bosonic theories with $\mathbb{Z}_{2}$ global symmetry.  These possible anomalies for such theories have a $\mathbb{Z}_{2}\times \mathbb{Z}_{2}$ classification and have recently been discussed in the condensed matter literature in \cite{fidkowski2019disentangling}.  We can characterize both such anomalies by inflow from a five-dimensional bulk $M_{5}$.  Let $a\in H^{1}(M,\mathbb{Z}_{2})$ denote the $\mathbb{Z}_{2}$ gauge field for the global symmetry.  Then the bulk SPTs are given by the following classical actions.
\begin{equation}\label{5dspt}
\exp\left(i \pi \int_{M_{5}}a\cup a\cup a\cup a\cup a\right)~, \hspace{.5in} \text{and}\hspace{.5in}\exp\left(i \pi \int_{M_{5}}a\cup w_{2}\cup w_{2}\right)~,
\end{equation}
where above $w_{2}\in H^{2}(M_{5},\mathbb{Z}_{2})$ is the second Stiefel-Whitney class of $M_{5}$.  Mathematically, the first SPT above, is captured by the group cohomology $H^{5}(\mathbb{Z}_{2},U(1))\cong \mathbb{Z}_{2}$, while the second involves the geometry non-trivially and goes beyond group cohomology.

We now consider $(2+1)d$ bosonic theories protected by time-reversal symmetry.  As expected from the general isomorphism \eqref{bossmith}, the anomalies of such theories also admit a $\mathbb{Z}_{2}\times \mathbb{Z}_{2}$ classification \cite{Kapustin:2014tfa}.  We can realize the $(3+1)d$ SPTs for these anomalies by performing a twisted compactification of the SPTs in \eqref{5dspt}.   Upon such a reduction, one factor of the gauge field $a$ is absorbed by the twisted circle direction, while the remaining factors each descend to the first Stiefel-Whitney class $w_{1}$ which can be viewed as a background gauge field for $\mathsf{T}$ symmetry (see Appendix \ref{Appsmith}).  This leads to
\begin{equation}\label{4dspt}
\exp\left(i \pi \int_{M_{4}}w_{1}\cup w_{1}\cup w_{1}\cup w_{1}\right)~, \hspace{.5in} \text{and}\hspace{.5in}\exp\left(i \pi \int_{M_{4}}w_{2}\cup w_{2}\right)~,
\end{equation}
which are indeed the correct $(2+1)d$ classical actions characterizing these SPTs.

\subsection{Discrete Gauge Theory Construction}

Let us give examples of $(3+1)d$ theories coupled to a $\mathbb{Z}_{2}$ global symmetry realizing these anomalies.  Consider for instance a version of $\mathbb{Z}_{2}$ gauge theory with dynamical gauge field $x_{1}\in H^{1}(M_{4}, \mathbb{Z}_{2})$ and magnetic dual $y_{2}\in H^{2}(M_{4}, \mathbb{Z}_{2})$.\footnote{In the terminology of \cite{Benini:2018reh, WIP}, $\mathbb{Z}_{2}$ gauge theory has an intrinsic 1-form symmetry $\mathbb{Z}_{2}^{(1)}$ and an intrinsic 2-form symmetry $\mathbb{Z}_{2}^{(2)}$.  If $A_{2}$ and $B_{3}$ are the associated background fields, the anomaly is $\exp\left(i\pi \int_{M_{5}}A_{2}\cup B_{3}\right)$.  Below we couple the $\mathbb{Z}_{2}$ 0-form global symmetry of interest to $\mathbb{Z}_{2}$ gauge theory through these higher-form symmetries.}  We couple this theory to the $\mathbb{Z}_{2}$ global symmetry through the action:
\begin{equation}\label{t1}
i \pi \int_{M_{4}} \left(x_{1} \cup \delta y_{2} +x_{1} \cup a\cup a \cup a+y_{2} \cup a\cup a\right)~.
\end{equation}
Physically, one can think of these couplings as implying that the worldvolume theory of the dynamical extended operators in the theory (Wilson lines and surface operators) carry various anomalies for the $\mathbb{Z}_{2}$ global symmetry.  For instance, $x_{1}$ has non-trivial holonomy surrounding a surface operator and the above leads to a cubic anomaly for the $\mathbb{Z}_{2}$ global symmetry inflowing onto this surface.  It is straightforward to compute that this theory realizes the  $a^{5}$ type anomaly described in \eqref{5dspt}.

Analogously, we can realize the $a w_{2}^{2}$ type anomaly by a version of $\mathbb{Z}_{2}$ gauge theory that is coupled to global symmetry as 
\begin{equation}\label{t2}
i \pi \int_{M_{4}} \left(x_{1} \cup \delta y_{2} +x_{1} \cup a\cup w_{2}+y_{2} \cup w_{2}\right)~.
\end{equation}
The couplings to $w_{2}$ now lead to worldvolume anomalies for the extended operators which involve their spin.  For instance, $y_{2}$ has a non-trivial integral on a sphere surrounding a Wilson line and the coupling above then implies that in this theory such a line is a fermion.

In both the theories \eqref{t1} and \eqref{t2} we can isolate the theory on the $\mathbb{Z}_{2}$ symmetry defect and exhibit the correspondence \eqref{bossmith}.  Upon twisted compactification, in each case we find a sector that is a version of a $(2+1)d$ $\mathbb{Z}_{2}$ gauge theory with dynamical fields $x_{1}$ and $y_{1}$ and modified quantum numbers for the line defects.  Specifically, twisted compactification of \eqref{t1} leads to 
\begin{equation}\label{t13d}
i \pi \int_{M_{3}} \left( x_{1} \cup \delta y_{1} +x_{1} \cup  w_{1}\cup  w_{1} +y_{1} \cup w_{1}\cup  w_{1}\right)~,
\end{equation}
which has both Wilson lines for $x_{1}$ and $y_{1}$ as Kramers doublets realizing the $w_{1}^{4}$ anomaly in \eqref{4dspt}.  Meanwhile twisted compactification of \eqref{t2} results in 
\begin{equation}\label{t23d}
i \pi \int_{M_{3}} \left(x_{1} \cup \delta y_{1} +x_{1} \cup  w_{2} +y_{1} \cup w_{2}\right)~.
\end{equation}
So in this case both Wilson lines for $x_{1}$ and $y_{1}$ are fermions living on the worldvolume of the bulk $\mathbb{Z}_{2}$ symmetry defect.

\section*{Acknowledgements}
We thank C.-M. Chang,  P.-S. Hsin, Z. Komargodski, M. Levin, Y.-H. Lin, A. Neitzke, S. Ryu, N. Seiberg, Y. Tachikawa, J. Wang, Y. Wang for interesting discussions. S.H.S. is supported by NSF grant PHY-1606531, the Roger Dashen Membership, the Simons
Foundation/SFARI (651444, NS). F.Y. is supported by DOE grant DE-SC0010008. This work benefited from the 2019 Pollica summer workshop, which was supported in part by the Simons Collaboration on the Non-Perturbative Bootstrap and in part by the INFN.

\appendix

\section{Smith Isomorphism Between Cobordism Groups}\label{Appsmith}
Here we summarize the mathematical formulation of the isomorphism \eqref{bossmith} between the $\mathbb{Z}_2$ SPT phases in $d$-dimensions and time-reversal SPTs in $d-1$-dimensions, following \cite{Kapustin:2014dxa, Tachikawa:2018njr}. For mathematics literature, see \cite{bahri1987eta,GilkeyBook}.

We start with the isomorphism between interacting bosonic SPT phases. The $d$-dimensional interacting bosonic SPT phases with a unitary $\bZ_2$ symmetry are classified by the following cobordism group \cite{Kapustin:2014tfa,Kapustin:2014dxa}:
\begin{equation}
	\Omega^d_{\text{SO}}(B\bZ_2):=\text{Hom}\left(\Omega_d^{\text{SO}}(B\bZ_2), U(1)\right),
\end{equation}
where $\Omega_d^{\text{SO}}(B\bZ_2)$ is the bordism group of pairs $(M,a)$ of a $d$-dimensional oriented manifold $M$ and a $\bZ_2$ background $a\in H^1(M,\bZ_2)$ on $M$.
The SPTs that can be made trivial after breaking the symmetry $G$ are classified by the reduced cobordism group
\begin{equation}
	\widetilde{\Omega}^d_{\text{SO}}(B\bZ_2):=\text{Hom}\left(\widetilde{\Omega}_d^{\text{SO}}(B\bZ_2), U(1)\right),
\end{equation}
where the reduced bordism group $\widetilde{\Omega}_{\text{SO}}^d(\bZ_2)$ is the quotient $\Omega_{\text{SO}}^d(\bZ_2)/\Omega_{\text{SO}}^d(\mathrm{pt})$.

On the other hand, $(d-1)$-dimensional interacting fermionic SPT phases with a time-reversal symmetry $\mathsf{T}$ are classified by 
\begin{equation}
\Omega^{d-1}_{\text{O}}(\text{pt}):=\text{Hom}\big(\Omega_{d-1}^{\text{O}}(\text{pt}), U(1)\big),
\end{equation}
where $\Omega_{d-1}^{\text{O}}(\text{pt})$ is the bordism group of $(d-1)$-dimensional unoriented manifolds to a point. 

The isomorphism \eqref{bossmith} relates the two bordism groups:
\begin{equation}
	f:\widetilde{\Omega}_d^{\text{SO}}(B\Z_2) \stackrel{\sim}{\rightarrow} \Omega_{d-1}^{\text{O}}(\text{pt}).
	\label{eq:smith}
\end{equation}
Concretely, given a bordism class $[M,a]\in\widetilde{\Omega}_d^{\text{SO}}(B\Z_2)$ with a representative pair $(M,a)$, its image under the isomorphism $f$ is the bordism class $[Y]\in\Omega_{d-1}^{\text{O}}(\text{pt})$ where $Y \subset M$ is a submanifold of $M$ that is Poincar\'e dual to $a$.\footnote{The Poncar\'e dual of $a\in H^1(M,\bZ_2)$ can be taken to be a smooth submanifold. $a$ defines a classifying map $a:M \to \mathbb{RP}^k$ with sufficiently large $k$, and the Poincar\'e dual can be taken to be the inverse image of $\mathbb{RP}^{k-1}\subset \mathbb{RP}^k$ under $a$. As generic $a$ is transverse to $\mathbb{RP}^{k-1}$, $a^{-1}(\mathbb{RP}^{k-1})$ defines a smooth submanifold of $M$.}
The isomorphism $f$ induces the correspondence between SPTs with $\bZ_2$ symmetry and those with time-reversal symmetry:
\begin{equation}
	\label{eq:fstar}
	f^*:\Omega^{d-1}_{\text{O}}(\mathrm{pt}) \stackrel{\sim}{\rightarrow} \widetilde{\Omega}^{d}_{\text{SO}}(B\bZ_2).
\end{equation}
This direction of the isomorphism means that given a time-reversal SPT in $d-1$ dimensions, one can construct a $d$-dimensional $\bZ_2$ SPT by demanding that the $\bZ_2$ symmetry defect supports the given time-reversal SPT.
An explicit form of $f^*$ can be described as follows generalizing the map discussed in Section \ref{genres}.
The theorem in \cite{thom1954quelques, brown1964relations} asserts that any cobordism invarinat in $\Omega^{d-1}_{\text{O}}(\mathrm{pt})$ can be represented as an polyonial $P(w_i)$, wehre $w_i$ are the Stiefel Whitney classes. Similarly, a bordism invaiant in $\Omega^{d}_{\text{SO}}(B\mathbb{Z}_2)$ can be written as a polyonial $\widetilde{P}$ of $\bZ_2$ backgound $a$ and $w_{i\ge2}$. The map \eqref{eq:fstar} is
\begin{equation}
	f^*(P(w_1,w_{i\ge 2})) \mapsto \widetilde{P}(a,w_{i\ge 2}) = a\cup P(a,w_{i\ge 2}).
\end{equation}

The inverse of $f$
\begin{equation}
	g = f^{-1}:\widetilde{\Omega}_{d-1}^{\text{O}}(\mathrm{pt}) \stackrel{\sim}{\rightarrow} \widetilde{\Omega}_{d}^{\text{SO}}(B\bZ_2),
	\label{eq:smithinv}
\end{equation}
can be constructed as follows.
Given an unoriented $d-1$-dimensional manifold $Y$ representing a class of $\Omega_{d-1}^{\text{O}}(\mathrm{pt})$ with the orientation (real line) bundle $\mathrm{det}(TY)$, we construct the circle bundle $\widetilde{M}_Y$ by taking the fiber-wise unit circles of the direct sum $\mathrm{det}(TY)\oplus \underline{\mathbb{R}}$ of $\mathrm{det}(TY)$ and the trivial real line bundle $\underline{\mathbb{R}}$. Further, we take the quotient of $\widetilde{M}_Y$ by the $\bZ_2$ acting on the fiber as $\pi$ rotation and call it $M_Y$.
The map $g$ is set to be $g([Y]) = [(M_Y,a)]$, where $a$ is the $\mathbb{Z}_2$ connection corresponding to the $\bZ_2$ bundle $\widetilde{M}_Y\to M_Y$.
The manifold $M_Y$ is oriented because when the transition function between patches of $Y$ is orientation-reversing, it flips the orientation of the circle fiber by construction, so that it preserves the orientation of the total space.
The Poincar\'e dual of $a$ is by construction the base $Y$ of the bundle, and thus we have $f\circ g = \mathrm{id}$.

The Pontryagin dual $g^*$ of \eqref{eq:smithinv}:
\begin{equation}
	g^*:\widetilde{\Omega}^{d}_{\text{SO}}(\mathrm{pt}) \stackrel{\sim}{\rightarrow} \Omega^{d-1}_{\text{O}}(B\bZ_2)
	\label{eq:smithinvstar}
\end{equation}
describes the $\bZ_2$ twisted compactification of given $\bZ_2$ SPT on $M_Y$ to a time-reversal SPT on $Y$, which is the focus of the main text.
In this construction, the time-reversal action on $Y$ involves the flipping of the circle fiber.
The explicit form of $g^*$ is
\begin{equation}
	g^*(\widetilde{P}(a, w_{i\ge 2})) \mapsto P(w_1,w_{i\ge 2}) = \widetilde{P}(a+w_1,w_{i\ge 2})|_{a^1},
\end{equation}
where the right hand side means the sum of the coefficients of the terms in $\widetilde{P}(a+w_1,w_{i\ge 2})$ that are linear in $a$.

There is the generalization of the isomorphism between bosonic SPTs to the fermionic SPTs.\footnote{There are further generalized versions of homomorphisms among different kinds of bordisms, which are discussed in \cite{GilkeyBook,Kapustin:2014dxa,Tachikawa:2018njr}.}
Mathematically, the corresponding isomorphism between bordism groups is  
\begin{equation}
	\widetilde{\Omega}_d^{\text{spin}}(B\Z_2) \stackrel{\sim}{\rightarrow} \Omega_{d-1}^{\text{pin${}^{-}$}}(\text{pt}),
	\label{eq:smithspin}
\end{equation}
where $\Omega^{\text{spin}}_*$ and $\Omega^{\text{pin${}^-$}}_*$ represent the cobordism groups of spin- and pin${}^-$-manifolds. Pin${}^-$-structure corresponds to the time-reversal symmetry $T$ with $T^2=1$.
Given a spin-manifold $M$ with a $\bZ_2$ background $a$, we take the image of the manifolds as the Poincar\'e dual $Y$ of $a$ as before. The spin-structure of $M$ restricts onto the bundle $TM|_Y \cong TY \oplus NY$, where $NY$ is the normal bundle which is isomorphic to $\mathrm{det}(TY)$.\footnote{As $TM|_Y$ is oriented, we can take a local orthogonal frame so that the determinant of the transition function is 1. Then $NY \cong \mathrm{det}(TY)$ follows from the decomposition $TM|_Y\cong TY \oplus NY$.} Now, a pin${}^-$-structure on $Y$ is equivalent to a spin structure on $TY\oplus \mathrm{det}(TY)$, which can naturally be induced from the spin-structure on $M$.
Conversely, given a pin${}^-$-structure on $Y$, it naturally extends to a spin structure on $\mathrm{det}(TY)\oplus \underline{\mathbb{R}}$ and then can be restricted to a spin structure on $M_Y$.
Therefore, the maps \eqref{eq:smith} and \eqref{eq:smithinv} are generalized to the case of (s)pin bordisms \eqref{eq:smithspin}.

\bibliographystyle{utphys}
\bibliography{Z2Time}

\end{document}